\documentclass[%
 aip,
 amsmath,amssymb,
 reprint,%
]{revtex4-1}

\usepackage{graphicx}
\usepackage{dcolumn}
\usepackage{bm}
\usepackage{float}

\usepackage[utf8]{inputenc}
\usepackage[T1]{fontenc}
\usepackage{mathptmx}
\usepackage{etoolbox}
\usepackage[normalem]{ulem}

\def\d{{\rm d}}
\def\i{{\rm i}}

\usepackage{color}
\newcommand{\ks}{\textcolor{black}}

\makeatletter
\def\@email#1#2{%
 \endgroup
 \patchcmd{\titleblock@produce}
  {\frontmatter@RRAPformat}
  {\frontmatter@RRAPformat{\produce@RRAP{*#1\href{mailto:#2}{#2}}}\frontmatter@RRAPformat}
  {}{}
}%
\makeatother
\begin{document}

\preprint{AIP/123-QED}

\title{Mixed convection instability in a viscosity stratified flow in a vertical channel}
\author{Ankush}
 \affiliation{Department of Mathematics, Indian Institute of Technology Hyderabad, Kandi - 502 284, Telangana, India}
\author{P. A. L. Narayana}
 \affiliation{Department of Mathematics, Indian Institute of Technology Hyderabad, Kandi - 502 284, Telangana, India}
 \author{Kirti Chandra Sahu$^*$}
 \affiliation{Department of Chemical Engineering, Indian Institute of Technology Hyderabad, Kandi - 502 284, Telangana, India}

\email{ananth@math.iith.ac.in, ksahu@che.iith.ac.in }

\date{\today}

\begin{abstract}
The present study examines the linear instability characteristics of double-diffusive mixed convective flow in a vertical channel with viscosity stratification. The viscosity of the fluid is modelled as an exponential function of temperature and concentration, with an activation energy parameter determining its sensitivity to temperature variation. Three scenarios are considered: buoyancy force due to thermal diffusion only, buoyancy force due to temperature and solute acting in the same direction, and buoyancy force due to temperature and solute acting in opposite directions. A generalized eigenvalue problem is derived and solved numerically for linear stability analysis via the Chebyshev spectral collocation method. Results indicate that higher values of the activation energy parameter lead to increased flow stability. Additionally, when both buoyant forces act in opposite directions, the Schmidt number has both stabilizing and destabilizing effects across the range of activation energy parameters, similar to the case of pure thermal diffusion. Furthermore, the solutal-buoyancy-opposed base flow is found to be the most stable, while the solutal-buoyancy-assisted base flow is the least stable. As expected, an increase in Reynolds number is shown to decrease the critical Rayleigh number.
\end{abstract}

\maketitle

\section{Introduction}
\label{sec:intro}

Flow instabilities driven by viscosity stratification due to concentration and temperature gradients are common in industrial applications and natural phenomena \cite{joseph1997core,selvam2007stability,govindarajan2014instabilities,chen1996miscible,petitjeans1996miscible,sahu2009linear}. The instability resulting from the interplay of varying temperature and solute concentration is termed thermo-solutal mixed convection. This type of convection is frequently encountered in many practical applications, including the transportation of crude oil through pipelines \cite{joseph1997core}, polymer processing \cite{pearson1985mechanics}, the chemical process industry \cite{cao2003instability}, and food and beverages processing \cite{regner2007predicting}, to name a few. \ks{Specifically, in biological and mechanical engineering applications, the flow dynamics due to the concentration and temperature gradients along the channel walls have been investigated by \citet{williams2020diffusioosmotic} and \citet{hu2021nanofluid}, respectively.} Moreover, in various engineering applications, including nuclear reactors, heat exchangers, electronic equipment, petroleum recovery, food processing, and biomedical devices, both temperature and concentration can exhibit variations along the boundary \cite{nazir2021applications,chen1996linear}. Thus, a fundamental understanding of the instabilities in thermo-solutal mixed convective flows can be helpful in many real-world applications. Although many researchers have investigated interfacial instability in immiscible fluids with viscosity contrast \cite{yih1967instability,mu2021interfacial}, in the following, we exclusively focus on the miscible configuration, which is considered in the present study. 

Several researchers have employed linear stability analysis to investigate the instabilities in viscosity-stratified shear flows caused by temperature gradients. In non-isothermal channel flow, while \citet{potter1972stability,pinarbasi1995role} demonstrated that the temperature difference between the walls always destabilizes the flow, \citet{wall1996linear, sameen2007effect} found that the temperature difference between the walls stabilizes the flow. They employed the viscosity of the fluid at the hot wall and the average viscosity across the channel as their viscosity scales. However, they did not take into account the effect of viscous heating (also known as viscous dissipation), which was investigated by other researchers in channel \cite{sahu2010stability} and Couette \cite{yueh1996linear,sukanek1973stability} flows. In Couette flows, the viscous heating stabilizes the flow because of the coupling between velocity perturbations and the base state temperature gradient, which results in spatially inhomogeneous temperature fluctuations and lowers local viscosity and dissipation energy of the disturbances \cite{thomas2003influence}. On the other hand, in a channel flow, \citet{sahu2010stability} demonstrated that the viscous heating could be destabilizing. An energy budget analysis was conducted to explain the underlying physics at the onset of instability. The effect of temperature-dependent viscosity on Rayleigh-B{\'e}nard convection has also been studied \cite{booker1976thermal,booker1978further,stengel1982onset}. \citet{booker1976thermal} experimentally investigates the onset of convection at a high Prandtl number. It was found that the heat transport by convection decreases significantly as the ratio of the viscosities at the top and bottom boundaries is increased. \citet{booker1978further} observed that increasing the viscosity ratio at the top and bottom boundaries increases the critical Rayleigh number for instability. The increase in the critical Rayleigh number justifies the decrease in convective heat transfer. \citet{stengel1982onset} investigated how the temperature-dependent viscosity would affect the linear stability analysis of Rayleigh-B{\'e}nard convection. They found that the critical Rayleigh number is nearly constant for low viscosity ratios, increases at intermediate ratio values, approaches its maximum value at about the viscosity ratios of 3000, and then decreases. By conducting a linear stability analysis for viscosity-stratified thermal convection, \citet{thangam1986stability} showed that the fluid with variable viscosity is less stable than the fluid with constant viscosity when the mean Prandtl number exceeds 100.

A few researchers also investigated the instability brought on by the solutes that result in viscosity stratification in a channel flow \cite{sahu2009linear,ghosh2014linear,chattopadhyay2017core,pramanik2013linear,ranganathan2001stabilization}. \citet{ranganathan2001stabilization} demonstrated that laminar flow becomes unstable when the fluid near the wall is more viscous than the fluid at the centre of the channel by conducting a linear stability analysis. At low Reynolds numbers and high diffusivities, when the critical layer (the region where the axial velocity equals the phase speed of the dominant mode) overlaps with the mixed layer of varying viscosity, a new mode of instability in addition to the Tollmien-Schlichting mode was observed. On the other hand, when the less viscous fluid is positioned at the near wall region, a significant stabilization takes place \cite{govindarajan2004effect}. A linear stability analysis performed by \citet{sahu2009linear} revealed that the flow develops into a more catastrophic absolute instability for high viscosity ratios and low diffusivity values, which in turn causes the flow to migrate towards a transitional state via a nonlinear mechanism. Subsequently, \citet{ghosh2014linear,chattopadhyay2017core} and \citet{pramanik2013linear} extended this study to porous media flows considering velocity slip and Korteweg stresses, respectively. An extensive literature review on this topic can be found in \citet{govindarajan2014instabilities}.

All of the aforementioned studies considered the stratification in viscosity induced by temperature variations or due to the presence of a solute (single-component or SC system). In reality, however,  viscosity stratification can happen when temperature, a species, or perhaps many species are active simultaneously. When two species having different diffusivities are present in a system, the situation is known as a double-diffusive (DD) phenomenon. These species may cause stratifications in density \cite{turner1974double} or viscosity \cite{sahu2014review,govindarajan2014instabilities,sahu2020linear}. In the present study, we limit our discussion to the instability resulting from the double-diffusive effect in viscosity-stratified flows of two miscible fluids with uniform density throughout the flow. Double-diffusive convections are known to exhibit contour-intuitive effects in contrast to SC systems. \citet{sahu2011linear} conducted a linear stability study for a three-layer channel flow with viscosity decreasing towards the wall (a stable configuration in the context of SC flow) and demonstrated the existence of an unstable mode at low Reynolds numbers that is distinct from the Tollmien-Schlichting wave. The double-diffusive effect drives this unstable mode. Further, they found that, in the presence of the DD effect, the flow becomes absolutely unstable, as opposed to being merely mildly convectively unstable in the corresponding SC system having the same viscosity variation \cite{sahu2012spatio}. Subsequently, several researchers have also observed the DD instabilities in other flow configurations, e.g. displacement of a highly viscous fluid by a less viscous one in porous media \cite{swernath2007viscous,mishra2010influence}, Hele-Shaw cell \cite{pritchard2009linear,bratsun2022effect} and pressure-driven flow in a channel \cite{mishra2012double}. Recently, \citet{verma2022radial,maharana2022effects,maharana2023stability} also investigated the instability driven by a different viscosity product resulting from a chemical reaction at the interfacial region between two miscible fluids.

The thermo-solutal mixed convection is a special case of a double-diffusive phenomenon. \citet{khandelwal2021instabilities} conducted a stability analysis for a pressure-driven vertical channel flow with thermo-solutal mixed convection for fluids without viscosity stratification to examine the effect of buoyancy ratio. They found that as the diffusivity decreases, the stability of the flow decreases when the buoyant force from species diffusion occurs in the same direction as the buoyant thermal force. The present work is an extension of \citet{khandelwal2021instabilities} to incorporate the effect of viscosity stratification. We investigate the linear instability of thermo-solutal mixed convection flow with viscosity stratification in a vertical channel that has not been studied yet to the best of our knowledge. We consider viscosity as a function of temperature and concentration. Our study aims to investigate how viscosity stratification affects the instability in thermo-solutal mixed convection in a vertical channel.

The rest of this paper is structured as follows. Section \ref{sec:form} illustrates the mathematical formulation of the basic state and the linear disturbance equations. The numerical techniques and their validation are presented in Section \ref{sec:num}. The linear stability results are presented in Section \ref{sec:dis}. Finally,  we summarise the results in Section \ref{sec:conc}.

\section{Formulation}
\label{sec:form}

\begin{figure}[h!]
\centering
\includegraphics[scale=0.4]{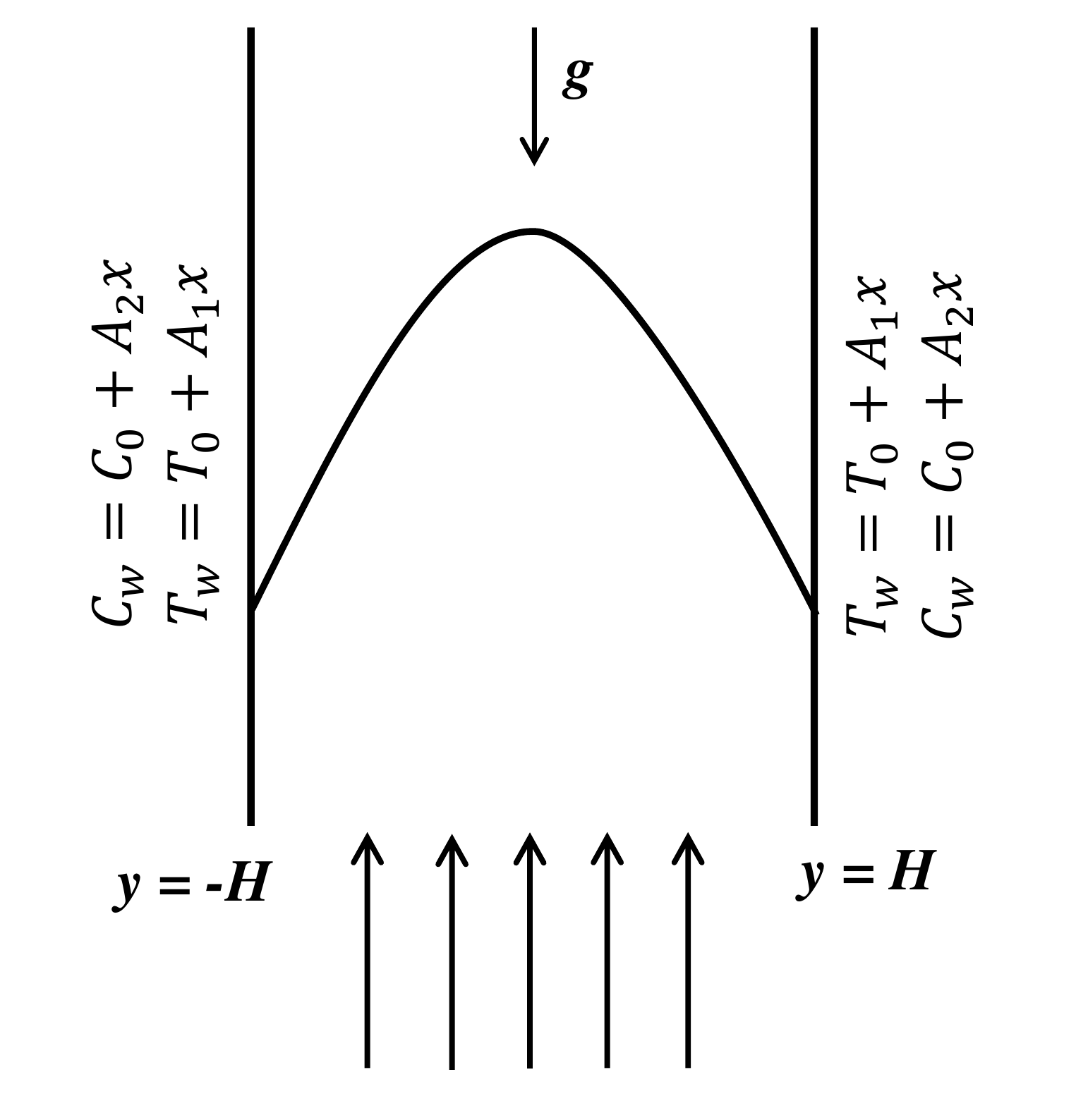}
\caption{Schematic diagram showing the thermo-solutal mixed convection flow with viscosity stratification in a vertical channel.}
\label{fig1:sch}
\end{figure} 

We investigate the linear stability characteristics of a pressure-driven thermo-solutal mixed convection flow of a Newtonian, incompressible, viscosity-stratified fluid in a vertical channel. A schematic diagram is shown in figure \ref{fig1:sch}. A Cartesian coordinate system $(x,y,z)$ is employed to formulate the problem, such that gravity acts in the negative $x$ direction. The channel walls are located at $y =\pm H$, wherein $H$ denotes the width of the half-channel. The channel walls are subjected to linear variations for temperature \cite{chen1996linear} and concentration along the $x$ direction, which are given by $T_w=T_0 +A_{1}x$ and $C_w=C_{0}+A_{2}x$. Here, $A_1$ and $A_2$ are constant temperature and concentration gradients;  $T_0$ and $C_0$ are the upstream reference temperature and solute concentration, respectively. Assuming that the temperature gradient is small, the variation in the density is small to be neglected everywhere except in the buoyancy term in the framework of Boussinesq’s approximation. This leads to  $\rho=\rho_f\left[1-\beta_{T}(T-T_w)-\beta_S(C-C_w)\right]$, where $\rho$, $\rho_f$, $T$, $T_w$, $C$, $C_w$, $\beta_T$ and $\beta_S$ are density, reference density, fluid temperature, wall temperature, instantaneous species concentration, concentration at the wall, volumetric thermal expansion coefficient and volumetric solute expansion coefficient, respectively. 

The dynamic viscosity $(\mu)$ varies with temperature and concentration, \ks{which is given by the Nahme-type viscosity-temperature relationship} \cite{nahme1940beitrage,sukanek1973stability}
\begin{equation}
    \mu =  \mu_r \exp\left [ \frac{\left ( C-C_{w} \right )}{A_{2}H Sc Re}  - \delta \frac{\left ( T-T_{w} \right )}{A_{1}HPrRe} \right ],
\label{eq1}
\end{equation}
where $\mu_r$ is the viscosity of the fluid at $T_0$ and $\delta$ is a dimensionless activation energy parameter that corresponds to the sensitivity of the viscosity to temperature variation.

The scalings employed used to nondimensionalise the governing equations are given by
\begin{equation}
\begin{split}
\left(\widetilde{x},\widetilde{y},\widetilde{z}\right)=\frac{1}{H}\left(x,y,z\right),~ \widetilde{t}=\frac{\overline{U}}{H}t, ~ \left(\widetilde{u},\widetilde{v},\widetilde{w}\right)=\frac{1}{\overline{U}}\left(u,v,w\right) \\ \widetilde{p}=\frac{p}{\rho_f\overline{U}^2}, ~ \widetilde{\mu}=\frac{\mu}{\mu_r}, ~ \theta=\frac{T-T_w}{A_1HRePr}, ~ \phi=\frac{C-C_w}{A_2HScRe}.
\label{eq2}
\end{split}
\end{equation}
Here,  $u$, $v$ and $w$ are the dimensional velocity components in the $x$, $y$ and $z$ directions, $\overline{U}$ is the average velocity, $t$ is dimensional time and $p$ is pressure. The corresponding dimensionless parameters are designated by superscript tilde notations. $\theta$ and $\phi$ are the dimensionless temperature and concentration, respectively. It is to be noted that the temperature and concentration are non-dimensionalised using the local temperature and concentration at the boundaries. This scaling results in the dimensionless temperature and concentration being zero at the boundaries. The dimensionless governing equations are given by 
\begin{eqnarray}
    {\partial u_{i} \over \partial x_{i}}=0,\label{eq3} \\
    {\partial u_{i} \over \partial t}+ u_{j} {\partial u_{i} \over \partial x_{j}}=-\frac{\partial p}{\partial x_{i}} &+& \nonumber \\ 
    \frac{1}{Re}\Big[\frac{\partial }{\partial x_{j}}\left\{\mu\left(\frac{\partial u_{i}}{\partial x_{j}}+\frac{\partial u_{j}}{\partial x_{i}}\right)\right\}\Big] &+&\frac{Ra}{Re}(\theta+N\phi)\delta_{1i}, \label{eq4} \\
    \frac{\partial \theta}{\partial t}+u_{j} {\partial \theta \over \partial x_{j}}=\frac{1}{RePr}\Big(\ks{{\partial^2 \theta \over \partial  x_{j} \partial  x_{j} }} &-&u_{j}\delta_{1j}\Big), \label{eq5} \\
    \frac{\partial \phi}{\partial t}+u_{j} {\partial \phi \over \partial x_{j}}=\frac{1}{ReSc}\Big(\ks{{\partial^2 \phi \over \partial  x_{j} \partial  x_{j} }} &-&u_{j}\delta_{1j}\Big), \label{eq6}
\end{eqnarray}
where $Re\left(\equiv \Bar{U}H/\nu\right)$, $Ra\left(\equiv g\beta_{T}A_{1}H^4/\nu k\right)$, $Pr\left(\equiv \nu/k\right) $, $Sc\left(\equiv \nu/D\right)$ and $N\left(\equiv\beta_{S}A_{2}k/\beta_{T} A_{1}D\right)$ are the Reynolds number, Rayleigh number, Prandtl number, Schmidt number, and buoyancy ratio, respectively; $\nu = \mu_r/\rho_f$ is the kinematic viscosity, $k$ is the thermal diffusivity and $D$ is the mass diffusivity. \ks{The derivation of equation (\ref{eq5}) is given in Appendix.}

\subsection{Basic state}
The linear stability characteristics of the flow is performed about an unperturbed, unidirectional, steady and fully-developed basic state profile. Under these assumptions, the above governing equations (\ref{eq3}-\ref{eq6}) are reduced to a set of ordinary differential equations, which are given by
\begin{eqnarray}
{\d \over \d y} \left(\mu_0 {\d U \over y}\right)+Ra(\Theta_0+N \Phi_0) &=& Re {\d P \over \d x},\label{eq7}
\\
{\d^2 \Theta_0 \over \d y^2} &=&U,\label{eq8}
\\
{\d^2\Phi_0 \over \d y^2}&=&U,\label{eq9}
\end{eqnarray}
where $U$,  $P$, $\Theta_0$,  $\Phi_0$  represent the velocity component in the $x$ direction, pressure, temperature, concentration, respectively. The dimensionless viscosity $(\mu_0)$ is given by
 \begin{equation}
 \mu_0=e^{\left(\Phi_0 - \delta\Theta_0\right)}.
 \end{equation}
Inspection of equation (\ref{eq7}) suggest that depending on the sign of $N$, the buoyancy caused by thermal diffusion can be designed to be either aligned with or opposed to the solutal buoyancy. The following boundary conditions are used to obtain the basic state profiles for $U$, $\Theta_0$ and $\Phi_0$.
\begin{equation}
U=\Theta_0=\Phi_0=0 \quad \text{at} \quad  y=\pm1.
\label{eq10}
\end{equation}
In addition, we impose a constant volumetric flow condition, which is given by $\int_{-1}^{1} U \,dy=2$.  The coupled equations (\ref{eq7}-\ref{eq9}) along with the boundary conditions [Eq. \ref{eq10}] are solved using MATLAB. We can also recover the base state equations presented in \citet{khandelwal2021instabilities} for a special case in our formulation by setting $\delta=1$.

\begin{figure}
\centering
\hspace{0.4cm} (a) \hspace{3.6cm} (b) \\
\includegraphics[width=0.22\textwidth]{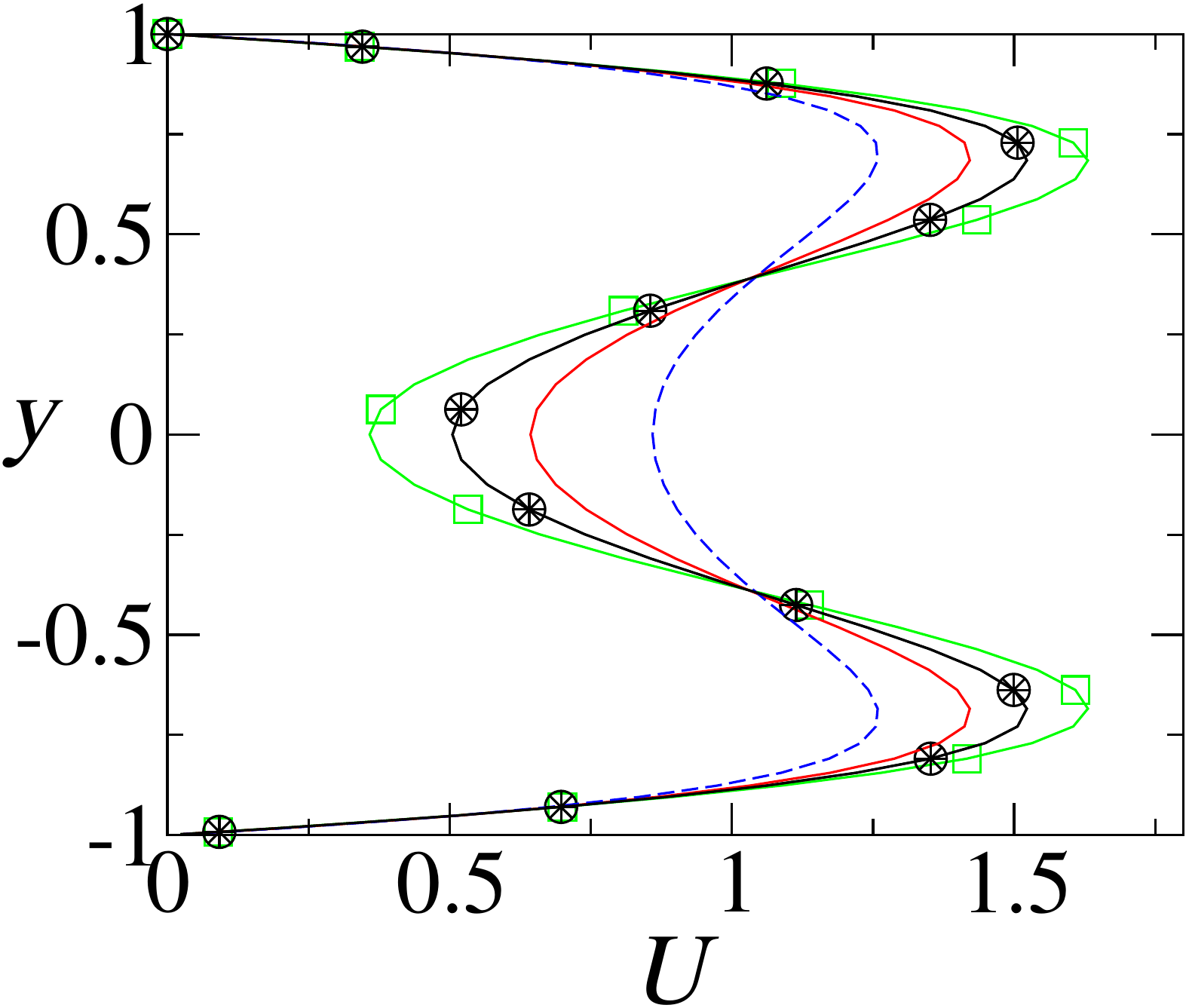} \hspace{2mm}
\includegraphics[width=0.22\textwidth]{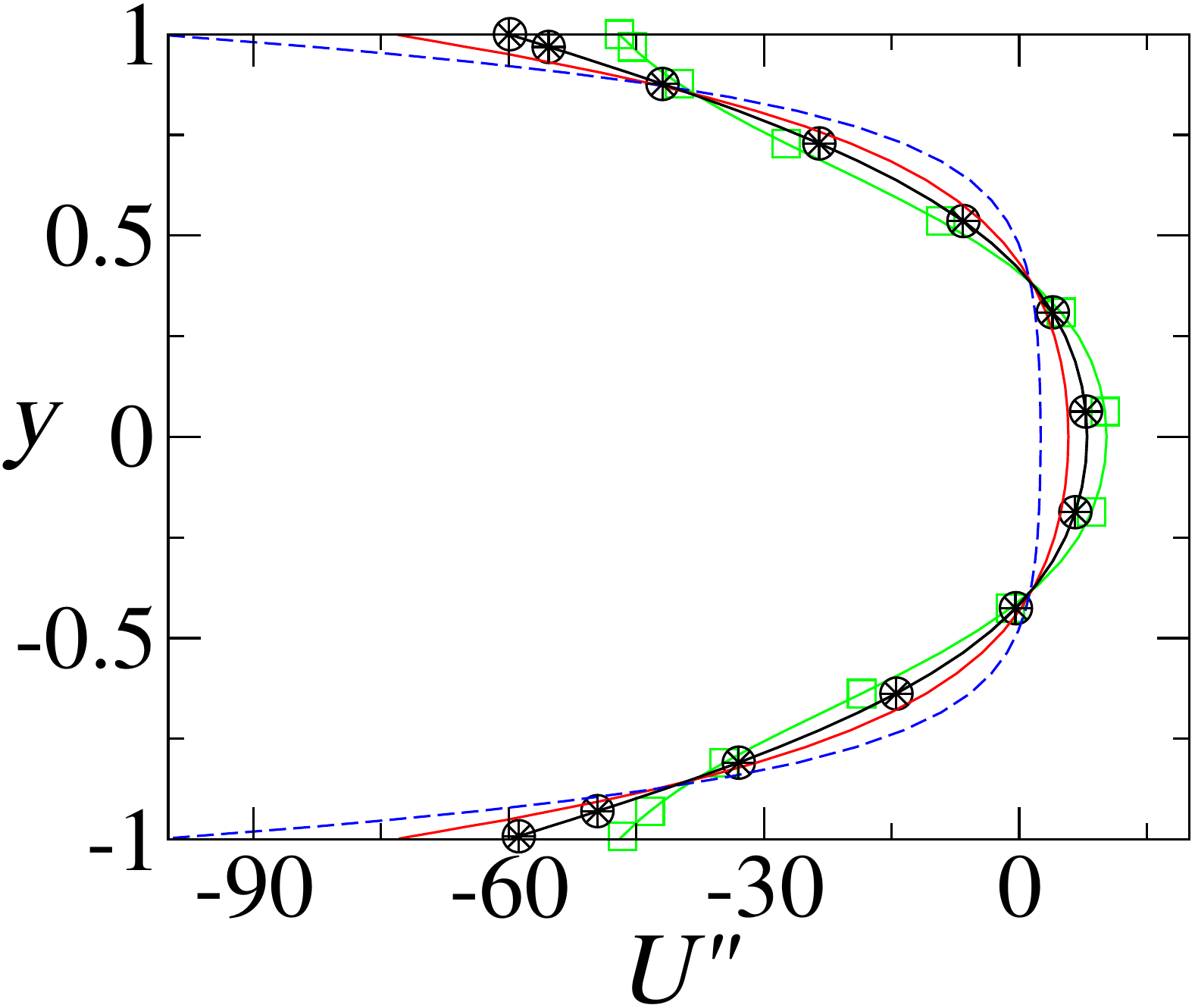} \\
\hspace{0.4cm} (c) \hspace{3.6cm} (d) \\
\includegraphics[width=0.22\textwidth]{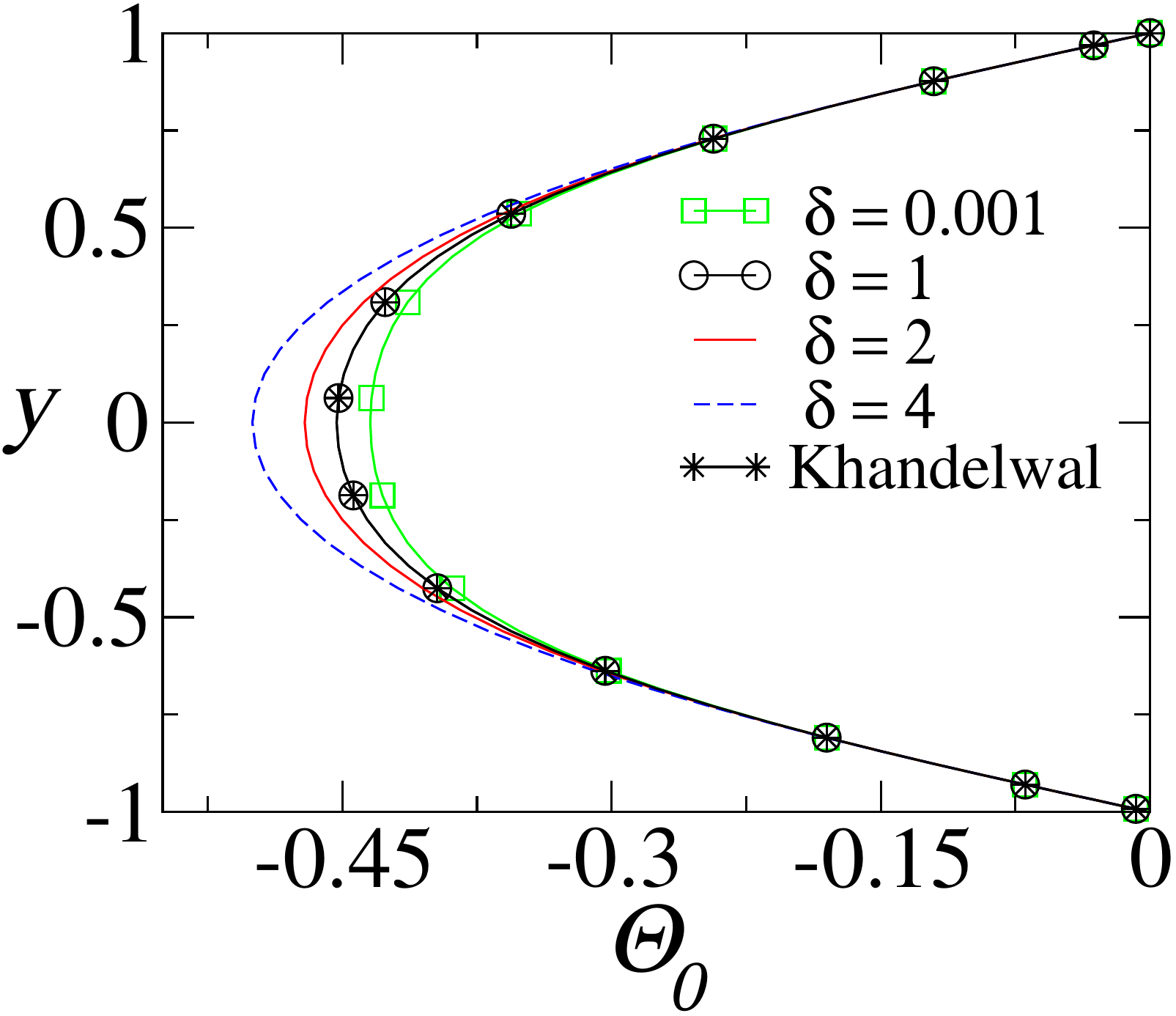}\hspace{2mm}
\includegraphics[width=0.22\textwidth]{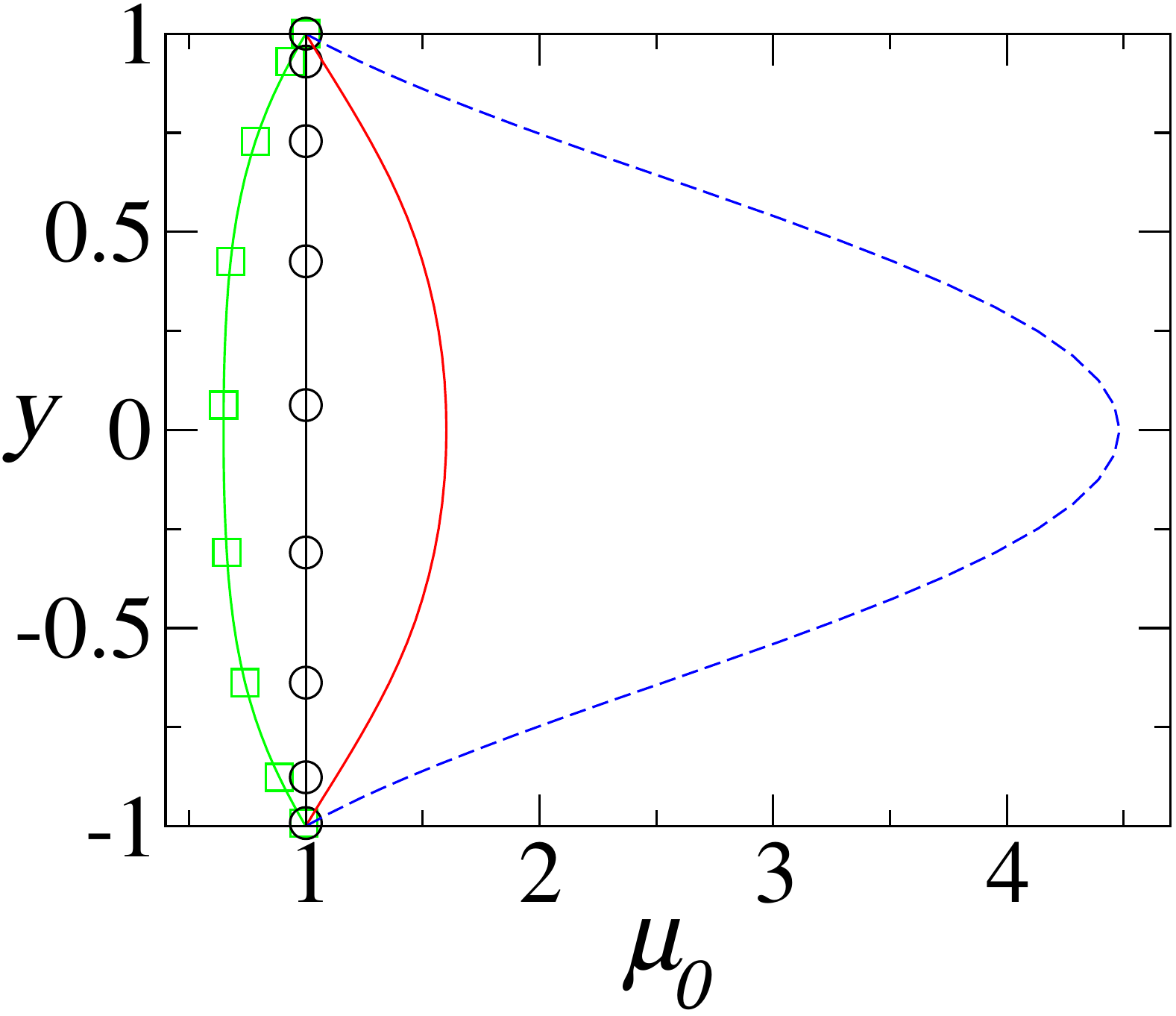}
\caption{Base state profiles of the (a) streamwise velocity $(U)$, (b) second derivative of velocity $(U^{\prime\prime})$, (c) temperature ($\Theta_0$) and (d) viscosity ($\mu_0$) for $N=0.5$ and $Ra=100$.}
\label{fig2:base}
\end{figure} 
Figure \ref{fig2:base}(a-d) depicts the base state profiles of velocity component $(U)$, its second order derivative $(U^{\prime\prime})$, temperature ($\Theta_0$) and viscosity ($\mu_0$) for different values of dimensionless activation energy parameter $\left(\delta \right)$ when $N=0.5$ and $Ra=100$. It can be observed in figure \ref{fig2:base}(a) and (b) that $U$ exhibits an inflectional profile for low values of $\delta$. The minimum value of $U$ at the centreline of the channel $(y=0)$ increases with increasing the value of $\delta$. The inflectional profile is a signature of Rayleigh inviscid instability \cite{rayleigh1879stability}. As expected, the temperature profile is negative throughout the domain with a minimum at the centreline and increasing the value of $\delta$ decreases the temperature (figure \ref{fig2:base}c). This, in turn, increases the viscosity of the fluid in the core region of the channel (figure \ref{fig2:base}d).

\subsection{Linear Stability Analysis}
In this section, we formulate the linear stability equations by expressing each flow variable as the sum of the base state and a 3D perturbation (denoted by a hat) as
\begin{eqnarray} 
 \left(u_t,v_t,w_t,p_t,\theta_t,\phi_t,\mu_t \right)\left(x,y,z,t\right)= \nonumber  \\ 
 \left[U(y),0,0,P(x),\Theta_0(y), \Phi_0(y), \mu_0(y)\right] + \nonumber  \\ 
  \left(\hat{u},\hat{v},\hat{w},\hat{p},\hat{\theta},\hat{\phi},\hat{\mu} \right)\left(y\right) \exp [{\i \left( \alpha x+\beta z-\alpha ct\right)}].
     \label{eq11}
\end{eqnarray}
Here, the subscript `$t$' represents the total of basic and perturbation variables. In eq. (\ref{eq11}), $\alpha$ and $\beta$ are real valued streamwise and spanwise wavenumbers, respectively, and $c=c_r+\i c_i$ is a complex wave speed. The sign of $c_i$ determines the temporal stability behavior of the given mode. The mode is unstable if $c_i >0$, stable if $c_i <0$, and neutrally stable if $c_i=0$. The following linear stability equations (after suppressing the hat notation) are obtained by inserting eq. (\ref{eq11}) into eqs. (\ref{eq3}-\ref{eq6}), then subtracting the base state equations, subsequently linearizing, and finally removing the pressure perturbation from the equations. The linear stability equations are given by 
\begin{align}
&-\frac{1}{Re}\left[\mu_0\left\{v'''' -2\left( \alpha^2+\beta^2\right)v'' + \left(\alpha^2 + \beta^2 \right)^2 v   \right\} \right. \nonumber\\ &\hspace{0.5cm}\left. + 2\mu_0^{'}\left\{ v''' -\left( \alpha^2+\beta^2\right)v'  \right\}  + \mu_0^{''}\left\{ v'' +\left(\alpha^2+\beta^2\right)v\right\} \right]\nonumber\\& \hspace{0.6cm}+\frac{\i \alpha}{Re}\left[U^\prime \left\{\mu''+ \left(\alpha^2 + \beta^2 \right) \mu \right\} + 2U^{''}\mu' + U^{'''}\mu   \right]\nonumber\\& \hspace{0.7cm}  + \frac{Ra}{Re} \i \alpha \left(\theta' + N\phi'\right)+ \i \alpha U\left\{v''-\left(\alpha^2 + \beta^2 \right)v\right\}\nonumber\\& \hspace{0.7cm} - \i \alpha U^{''}  v = \i \alpha c\left[v''-\left(\alpha^2 + \beta^2 \right)v\right],
\label{eq12}
\end{align}
\begin{align}  
& -\frac{1}{Re}\left[\mu_0\left\{ \eta''-\left(\alpha^2 + \beta^2 \right)\eta \right\} +  \mu_0^{'} \eta' + \i \beta U^{''} \mu \right.\nonumber\\
& \hspace{0.5cm}\left. + \i \beta U^{'}\mu'\right] + \i \beta U^{'}v + \i \alpha U\eta  -\i \beta\frac{Ra}{Re}\left(\theta + N\phi \right) \nonumber \\& \hspace{0.5cm} = \i \alpha c\eta,
\label{eq13}
\end{align}    
\begin{align}  
& -\frac{1}{RePr}\left[ \theta''- \left(\alpha^2 + \beta^2 \right)\theta\right] + \frac{1}{\left(\alpha^2 + \beta^2 \right)RePr}\left(\i \alpha v' - \i \beta\eta\right) \nonumber\\
& \hspace{0.5cm} + \Theta_0^{'}v +\i \alpha U\theta=\i \alpha c\theta,
\label{eq14}
\end{align}
\begin{align}
& -\frac{1}{ ReSc}\left[ \phi''- \left(\alpha^2 + \beta^2 \right)\phi\right] + \frac{1}{\left(\alpha^2 + \beta^2 \right)ReSc}\left(\i \alpha v' - \i \beta \eta\right) \nonumber\\
& \hspace{0.5cm} + \Phi_0^{'}v + \i \alpha U\phi= \i \alpha c\phi,
\label{eq15}
\end{align}
\ks{Here, the linearised perturbation of the viscosity} $(\mu)$ is given by
\begin{equation}
    \mu=\mu_0\left(\phi-\delta \theta\right).
    \label{eq16}
\end{equation}
where the prime denotes differentiation with respect to $y$, and $\eta$ is a normal component of vorticity, which is defined as, $\eta = \i\beta u - \i\alpha w$. The corresponding disturbance boundary conditions at the channel walls are given as 
 \begin{equation}  
 v=v^\prime=\eta=\theta=\phi=0 \quad \text{at} \quad  y=\pm1.
 \label{eq17}
 \end{equation}
Eqs. (\ref{eq12})-(\ref{eq15}) along with boundary conditions [Eq. \ref{eq17}] forms a generalized eigenvalue problem for a complex disturbance wave speed $(c)$. It is to be noted that for a special case with $\delta=1$ and $\theta=\phi$, the stability equations reduce to those of \citet{khandelwal2021instabilities}. 

\section{numerical techniques and validation}
\label{sec:num}

\begin{figure*}
    \centering
    \hspace{0.2cm} (a) \hspace{5.8cm} (b)  \\
    \includegraphics[width=0.3\textwidth]{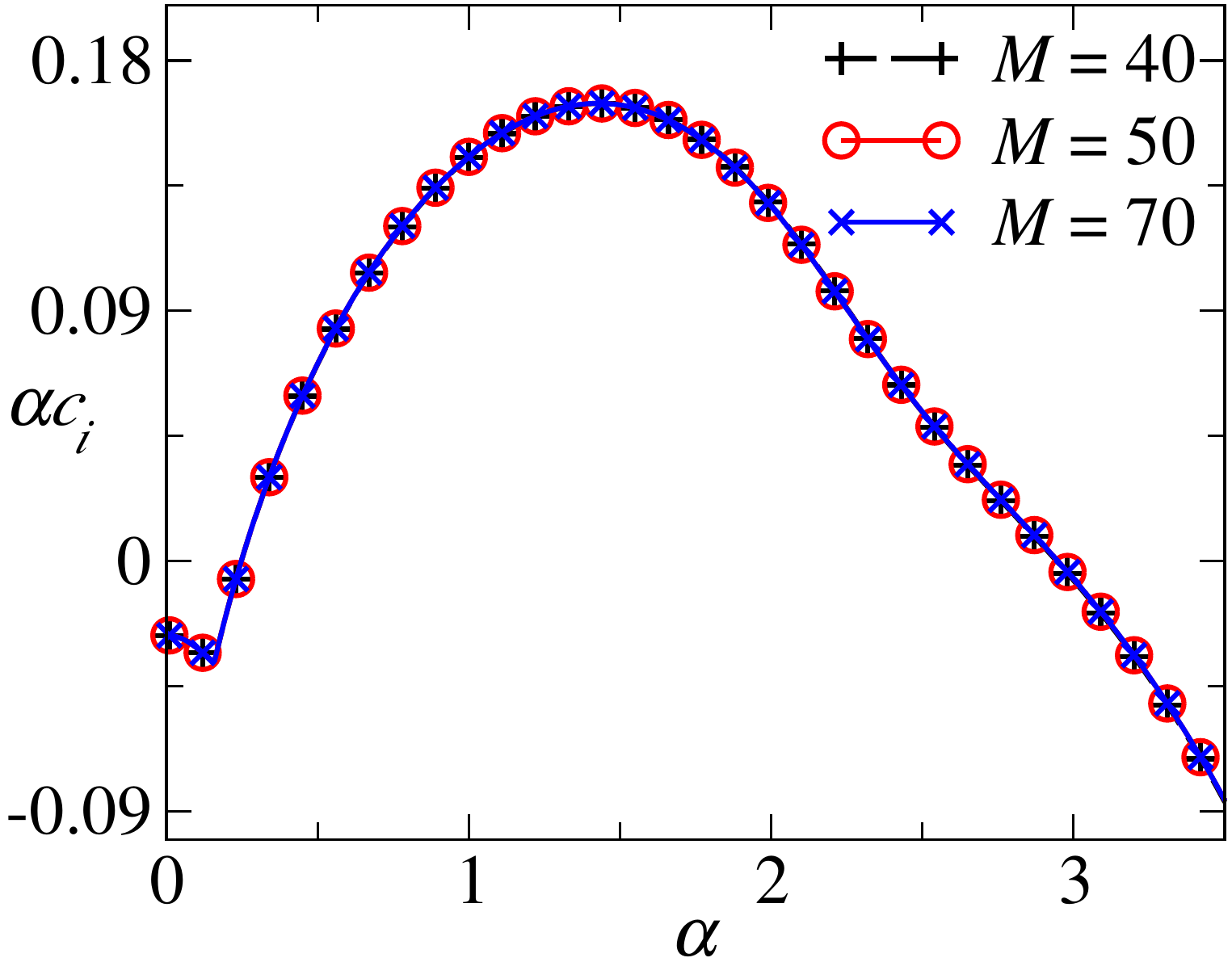}   \hspace{0.5cm}  \includegraphics[width=0.32\textwidth]{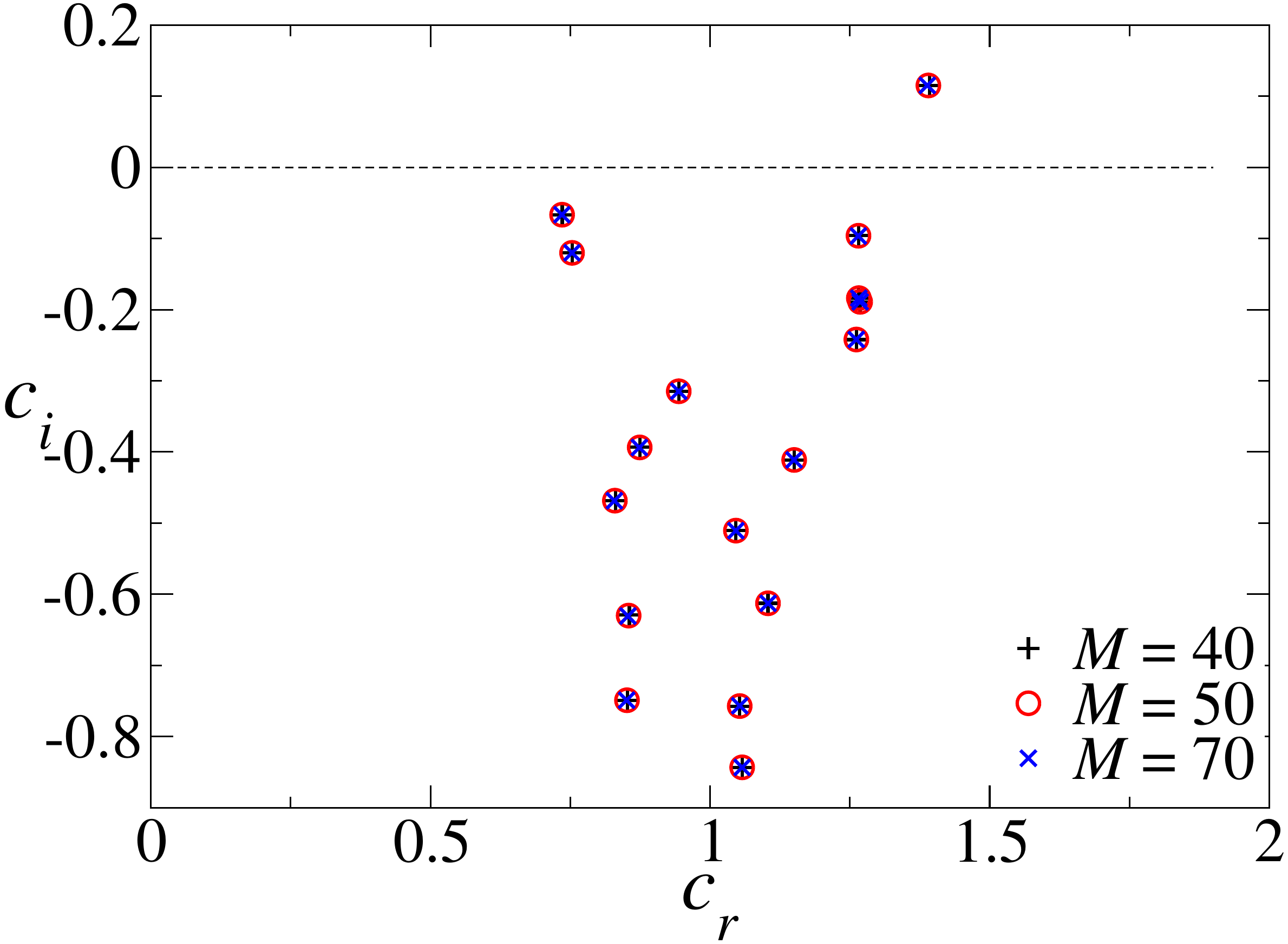} 
    \caption{Effect of the order of Chebyshev polynomials $(M)$ on the (a) growth rate $(\alpha c_i)$ of the disturbance and (b) eigenspectrum ($c_r$ versus $c_i$) of the most unstable mode for $\alpha=1.43$. The values of the remaining parameters are $Re=100$, $Ra=100$, $Pr=0.7, Sc=1$, $N=0.5$, $\beta=0$, and $\delta=2$.}
    \label{fig3}
\end{figure*} 
\begin{figure*}
\centering
\hspace{1cm} (a) \hspace{5.2cm} (b) \hspace{5.2cm} (c)  \\
\includegraphics[width=0.34\textwidth]{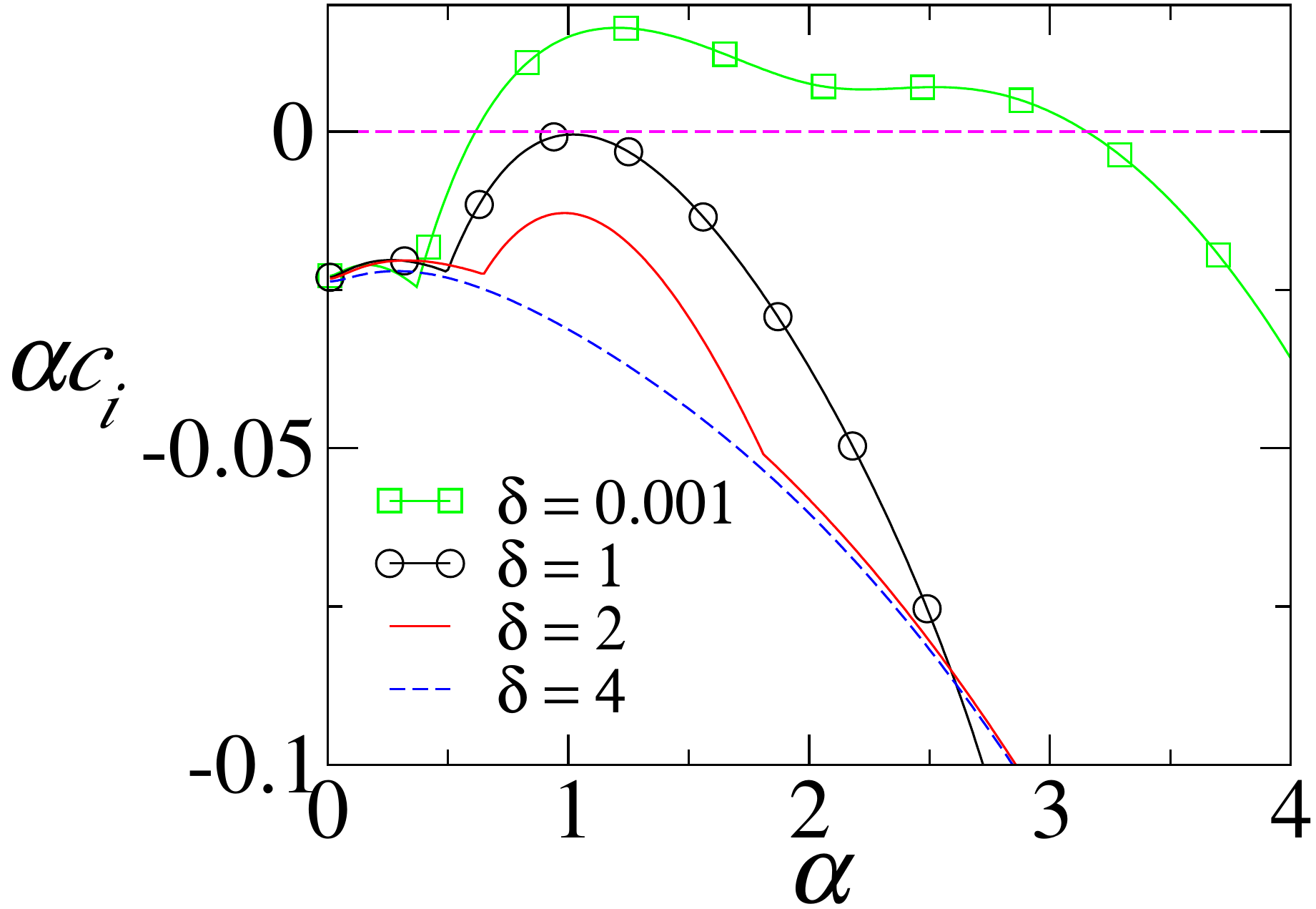}
\includegraphics[width=0.32\textwidth]{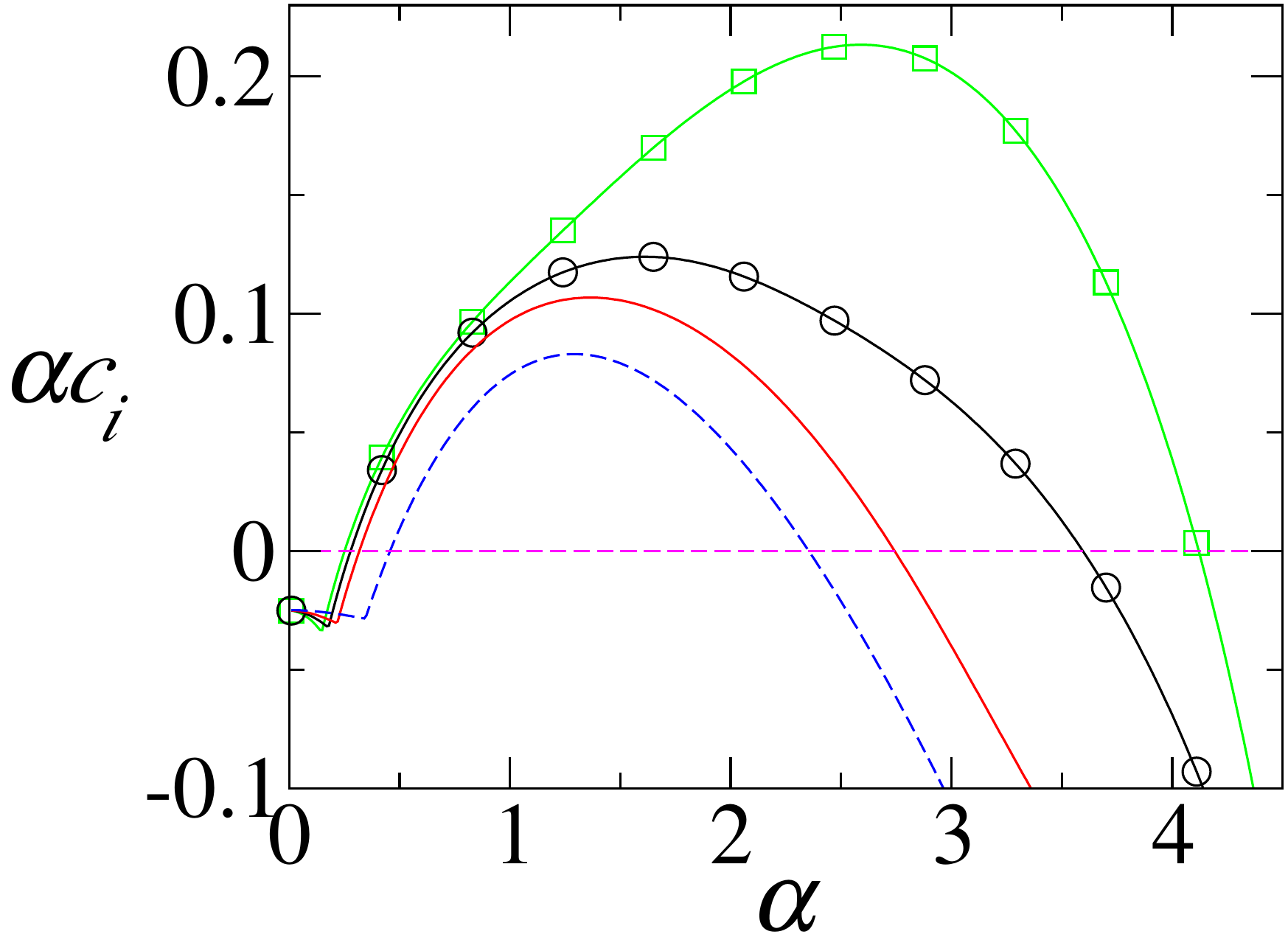}
\includegraphics[width=0.32\textwidth]{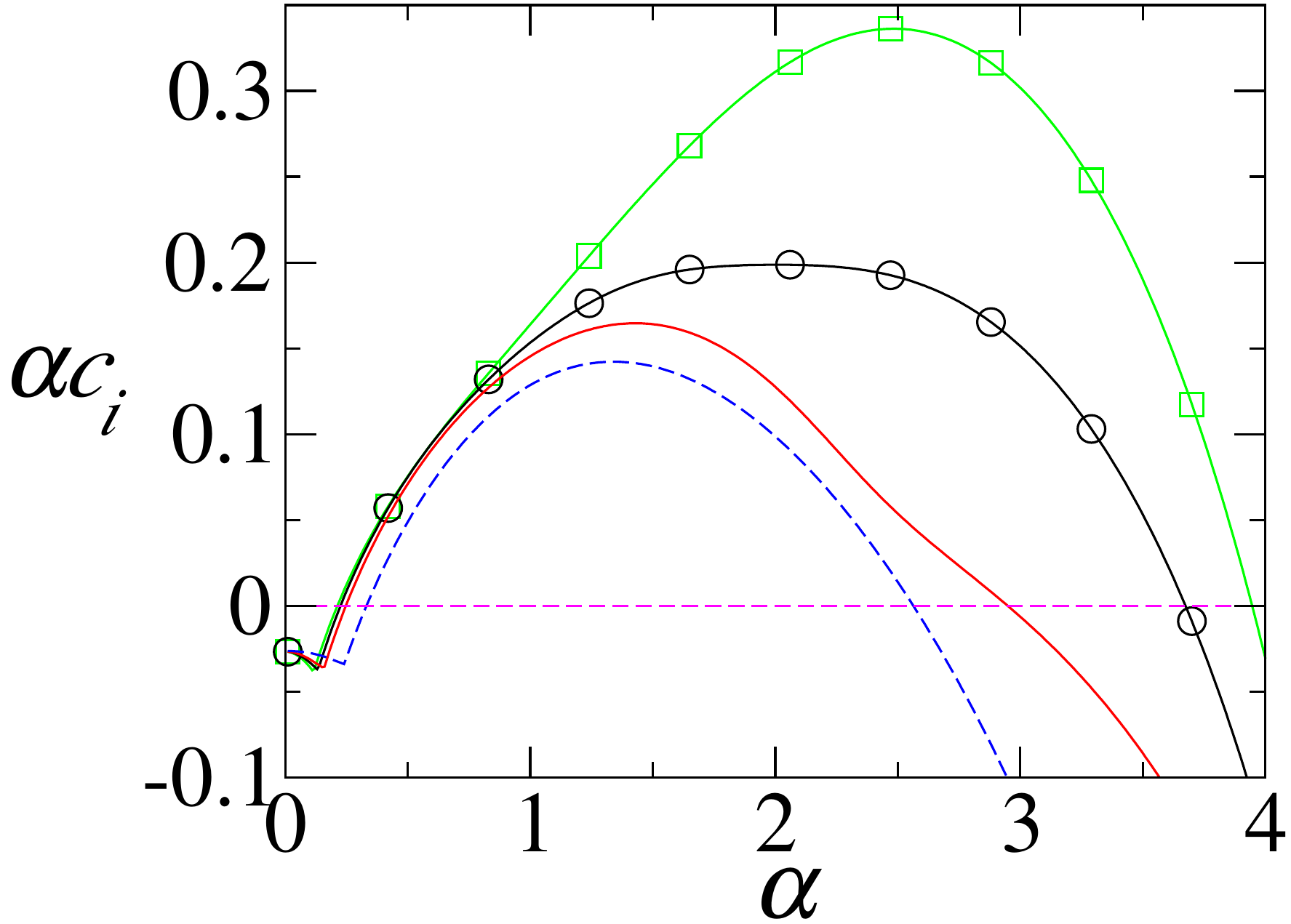}
\caption{Comparison of growth rate curves for different values of $\delta$. $(a)$ $N=-0.5$, $(b)$ $N=0$ and $(c)$ $N=0.5$. The values of the remaining parameters are $Pr=0.7$, $Re=100$, $Ra=100$  and $Sc=1$.}
\label{fig4}
\end{figure*}

We employ a Chebyshev spectral collocation method \cite{canuto1988spectral} to get the numerical solution of the linear stability equations with boundary conditions discussed in the previous section. The Gauss-Lobatto points are chosen as collocation points and they are given by
\begin{equation}
y_j=cos\left(\frac{\pi j}{M}\right), \quad j=0,1,2,...,M,
\label{eq18}
\end{equation}
where $M$ denotes for the order of the base polynomial, such that the $(M+1)$ number of grid points coincide with all the extremum of the Chebyshev polynomial $T_M(y)=\cos(M\cos^{-1}y)$ of order $M$. Upon discretization along the $y$-axis using the collocation points, the linear stability equations can be written as a generalized matrix eigenvalue problem, which is given by
\begin{align}
&\begin{bmatrix}
\mathcal{A}_{11} & \mathcal{A}_{12}  & \mathcal{A}_{13} & \mathcal{A}_{14}\\
\mathcal{A}_{21} & \mathcal{A}_{22}  & \mathcal{A}_{23} & \mathcal{A}_{24}\\
\mathcal{A}_{31} & \mathcal{A}_{32}  & \mathcal{A}_{33} & \mathcal{A}_{34}\\
\mathcal{A}_{41} & \mathcal{A}_{42}  & \mathcal{A}_{43} & \mathcal{A}_{44}\\
\end{bmatrix}
\begin{bmatrix}
v\\ \eta\\ \theta\\ \phi
\end{bmatrix}\nonumber \\& \hspace{0.5cm} =c
\begin{bmatrix}
\mathcal{B}_{11} & \mathcal{B}_{12}  & \mathcal{B}_{13} & \mathcal{B}_{14}\\
\mathcal{B}_{21} & \mathcal{B}_{22}  & \mathcal{B}_{23} & \mathcal{B}_{24}\\
\mathcal{B}_{31} & \mathcal{B}_{32}  & \mathcal{B}_{33} & \mathcal{B}_{34}\\
\mathcal{B}_{41} & \mathcal{B}_{42}  & \mathcal{B}_{43} & \mathcal{B}_{44}\\
\end{bmatrix}
\begin{bmatrix}
v\\ \eta\\ \theta\\ \phi
\end{bmatrix}.
\end{align}
In the above expression, an eigenvalue $c$ is determined using the MATLAB software.

To validate the numerical procedure, we examine the dependence of our numerical solution upon mesh refinement. This is done by comparing the growth rate curves ($\alpha c_i$ versus $\alpha$) obtained using different numbers of collocation points in figure \ref{fig3}(a). The values of the remaining parameters are $Re=100$, $Ra=100$, $Pr=0.7$, $Sc=1$, $N=0.5$, $\beta=0$, and $\delta=2$. It can be seen that the growth rate of the most unstable mode $(\alpha c_i)$ increases with increasing $\alpha$, reaches a maximum (most unstable wavelength) and then decreases to become negative at about $\alpha=3$ (cut-off wavelength). It can be seen that the growth rate curves are identical for different values of the order of Chebyshev polynomial $(M)$ indicating a numerically converged solution. Therefore, $M=50$ is fixed for the rest of the numerical simulations.
To validate the numerical procedure, we examine the dependence of our numerical solution upon mesh refinement. This is done by comparing the growth rate curves ($\alpha c_i$ versus $\alpha$) obtained using different numbers of collocation points in figure \ref{fig3}(a). Figure  \ref{fig3}(b) depicts the eigenspectrum ($c_r$ versus $c_i$) of the most unstable mode with $\alpha=1.43$ associated with figure \ref{fig3}(a). The values of the remaining parameters are $Re=100$, $Ra=100$, $Pr=0.7$, $Sc=1$, $N=0.5$, $\beta=0$ and $\delta=2$. It can be seen that the growth rate of the most unstable mode $(\alpha c_i)$ increases with increasing $\alpha$, reaches a maximum (most unstable wavelength) and then decreases to become negative at about $\alpha=3$ (cut-off wavelength). It can be seen that the growth rate curves and eigenspectrum of the most unstable mode are identical for different values of the order of Chebyshev polynomial $(M)$ indicating a numerically converged solution. Therefore, $M=50$ is fixed for the rest of the numerical simulations.

\section{Results and discussion}
\label{sec:dis}

We present the linear stability results for viscosity-stratified flow in a vertical channel affected by thermal-solutal mixed convection. The Reynolds number ($Re$), Rayleigh number ($Ra$), Prandtl number ($Pr$), buoyancy ratio ($N$), and Schmidt number ($Sc$) are the five independent dimensionless parameters that influence the stability characteristics of the flow in the configuration shown in figure \ref{fig1:sch}. The main goal of our investigation is to look at how viscosity variations affect the stability of base state flow in three different situations, namely when (i) total buoyant force is due to temperature and solute acting in the opposite directions (solutal-buoyancy-opposed flow, $N<0$), (ii) the buoyant force is only due to thermal diffusion (solutal-buoyancy neutral flow, $N=0$) and (iii) the total buoyant force is due to temperature and solute acting in the same directions (solutal-buoyancy-assisted flow, $N>0$). It is to be noted that we have incorporated the Squires's theorem, which states that for parallel shear flows, the two-dimensional perturbation $(\beta=0)$ is more unstable than the three-dimensional perturbation. \citet{khandelwal2021instabilities} also showed that two-dimensional disturbances are more dangerous than three-dimensional disturbances for a similar problem when there is no viscosity stratification. Thus, we restrict the analysis to streamwise wavenumber by setting $\beta = 0$ in our study.

As illustrated in figure \ref{fig4}(a)-(c), we begin by examining the growth rate of the disturbance $(\alpha c_i)$ in relation to the streamwise wavenumber $(\alpha)$ for various values of the activation energy parameter ($\delta$) for $N=-0.5$, 0 and 0.5, respectively. Note that $\delta=1$ corresponds to the constant viscosity case considered by \citet{khandelwal2021instabilities}. In figure \ref{fig4}(a)-(c), the values of the rest of the dimensionless parameters are $Re=100$, $Ra=100$, $Pr=0.7$ and $Sc=1$. The positive and negative values of the growth rate $(\alpha c_i)$ represent the situation when a given disturbance grows (unstable) or decays (stable) with time. Figure \ref{fig4}(a) (for $N=-0.5$) depicts that $\alpha c_i < 0$ for all values of $\alpha$ for $\delta=1$, $2$ and $4$. This indicates that the flow is stable for this set of parameters. In contrast, for $\delta=0.001$, $\alpha c_i > 0$ for $0.6 < \alpha < 3.14$ and negative for other values of $\alpha$. Thus, in the solutal-buoyancy-opposed flow configuration (with $N=-0.5$), the disturbances with wavenumbers $0.6 < \alpha < 3.14$ are unstable for $\delta=0.001$; the most unstable and the cut-off wavenumbers are $\alpha=1.2$ and 3.14, respectively. In the situations with $N=0$ (solutal-buoyancy neutral flow) and $N=0.5$ ((solutal-buoyancy-assisted flow), $\alpha c_i > 0$ for all values of $\delta$ considered in our study. It can be seen that in both these situations, the wavenumbers associated with the most-unstable and cut-off modes decrease with increasing the value of $\delta$. Thus, we can conclude that increasing $\delta$ has a stabilising influence. Close inspection of figure \ref{fig4}(a)-(c) also reveals that increasing $N$ has a destabilising influence for each value of $\delta$. For instance, the maximum growth rate $(\alpha c_i)$ increases as we increase the value of $N$, i.e. for $N=-0.5$, 0 and 0.5. 

\begin{figure*} 
\hspace{0.7cm} (a) \hspace{5.2cm} (b) \hspace{5.2cm} (c)  \\
\includegraphics[width=0.32\textwidth]{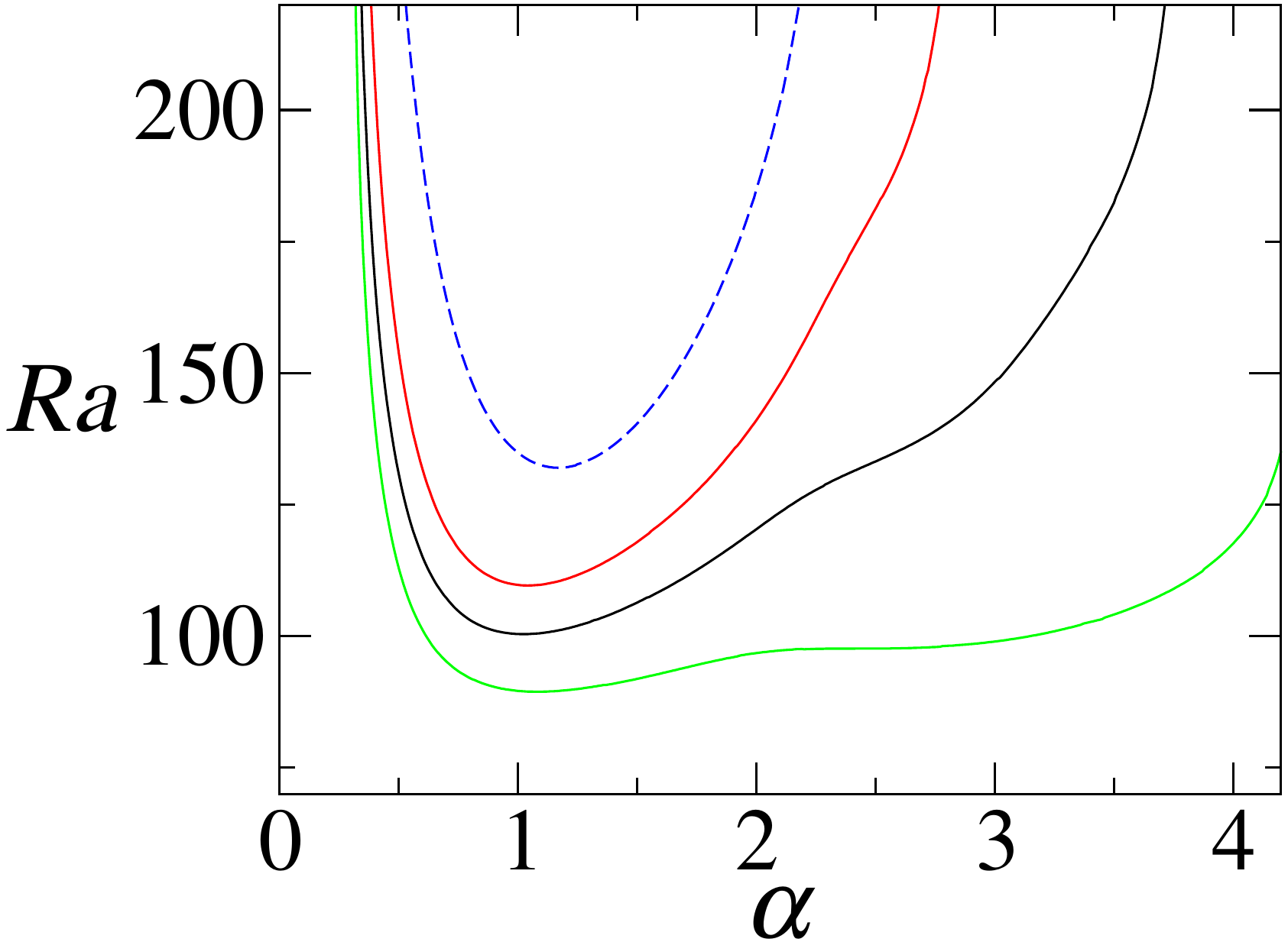}
\includegraphics[width=0.32\textwidth]{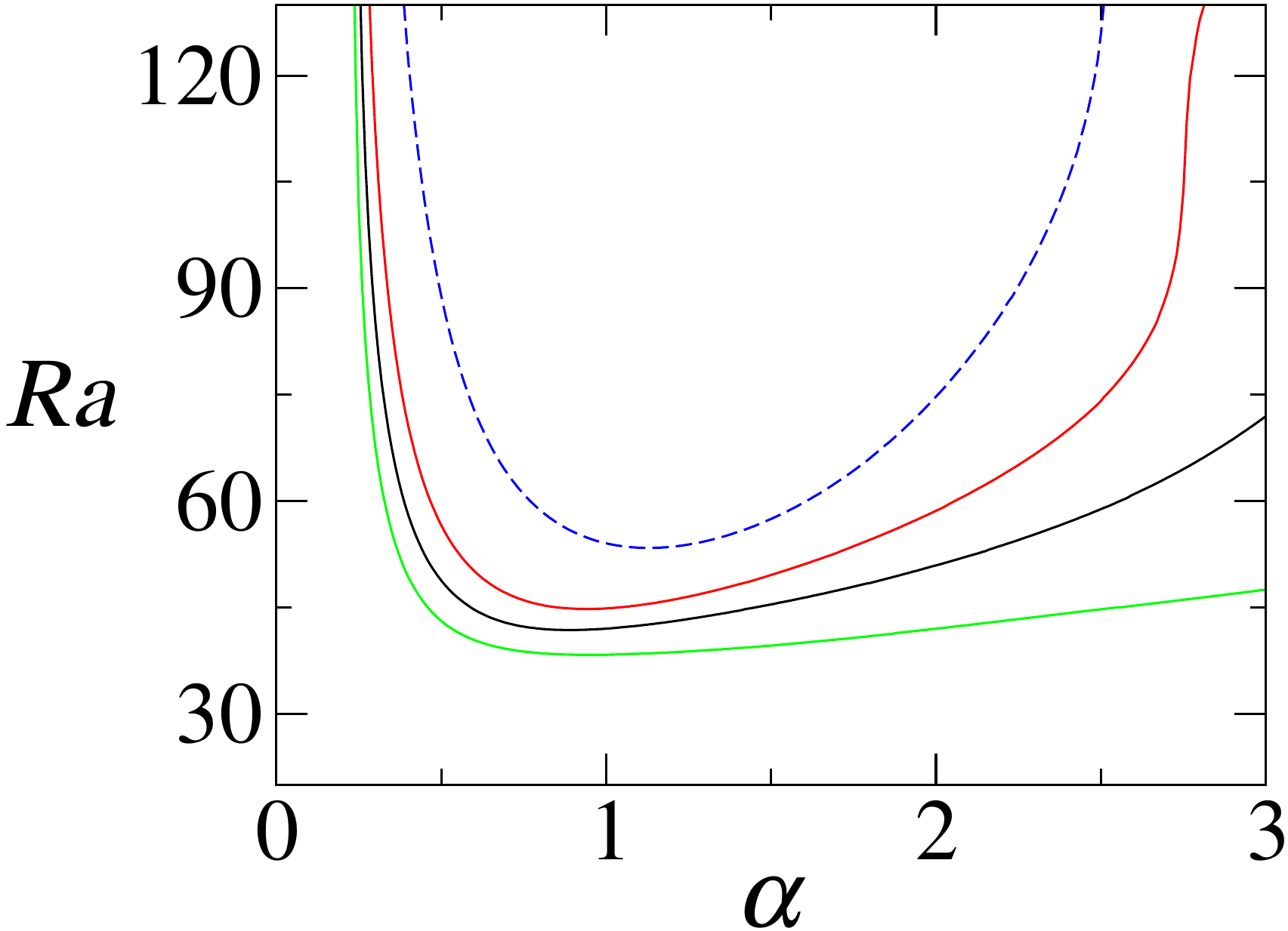}
\includegraphics[width=0.32\textwidth]{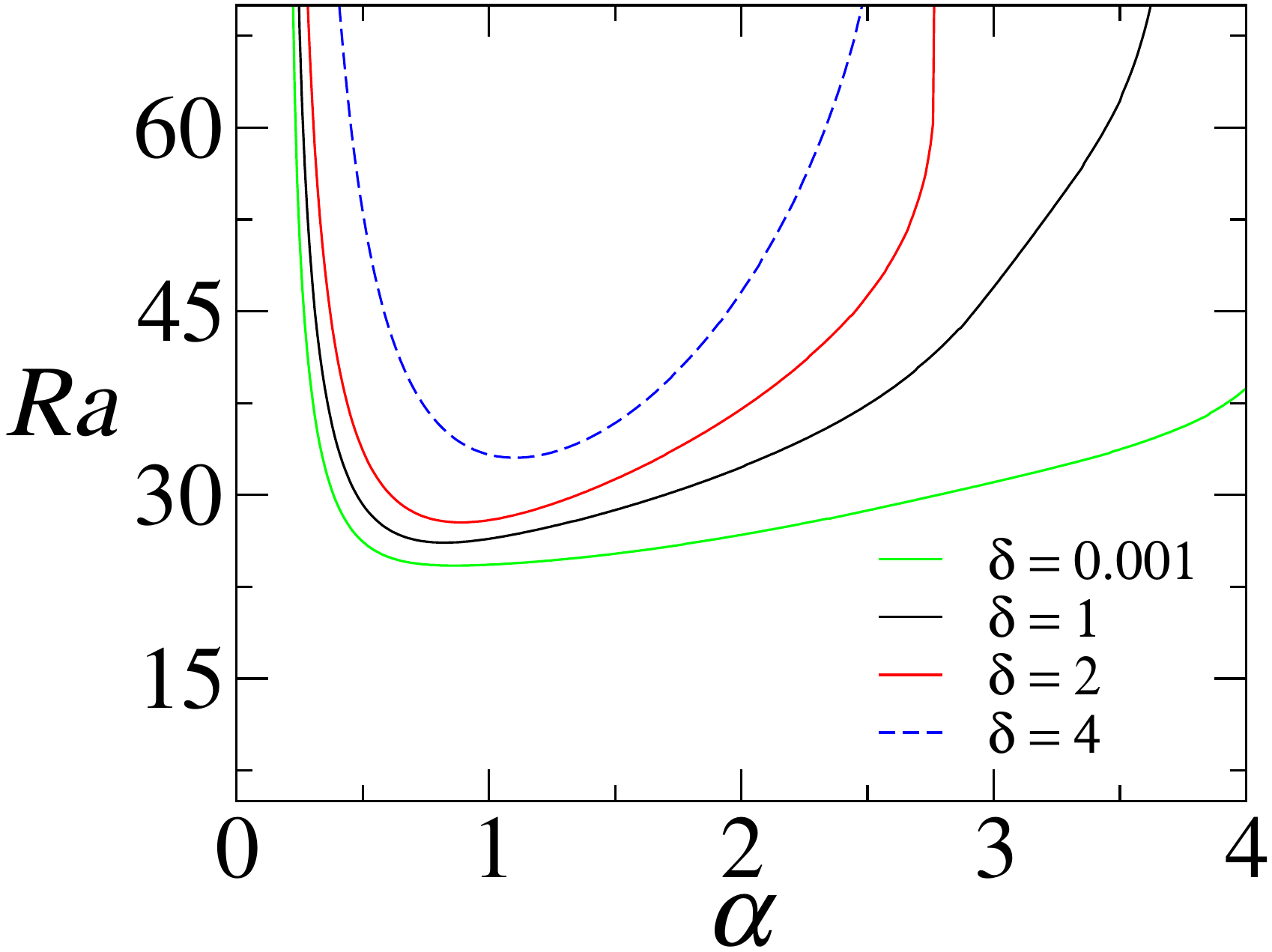}
\caption{Neutral stability curves for different values of $\delta$. (a) $N=-0.5$, (b) $N=0$ and (c) $N=0.5$.  The values of the remaining parameters are $Sc=1$, $Re=100$ and $Pr=0.7$.}
\label{fig5}
\end{figure*}

\begin{figure*} 
\centering
\hspace{0.7cm} (a) \hspace{5.2cm} (b) \hspace{5.2cm} (c)  \\
\includegraphics[width=0.32\textwidth]{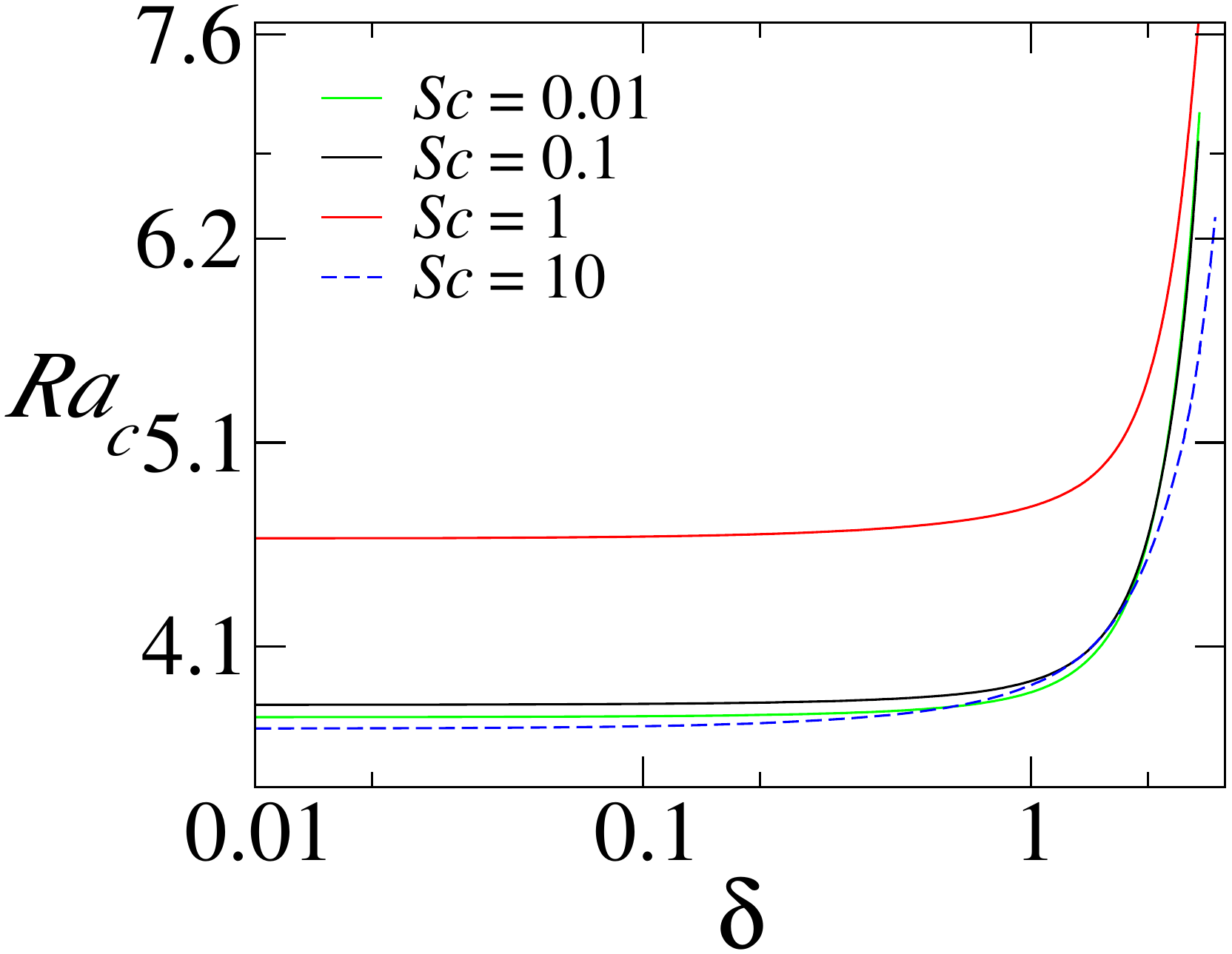}
\includegraphics[width=0.32\textwidth]{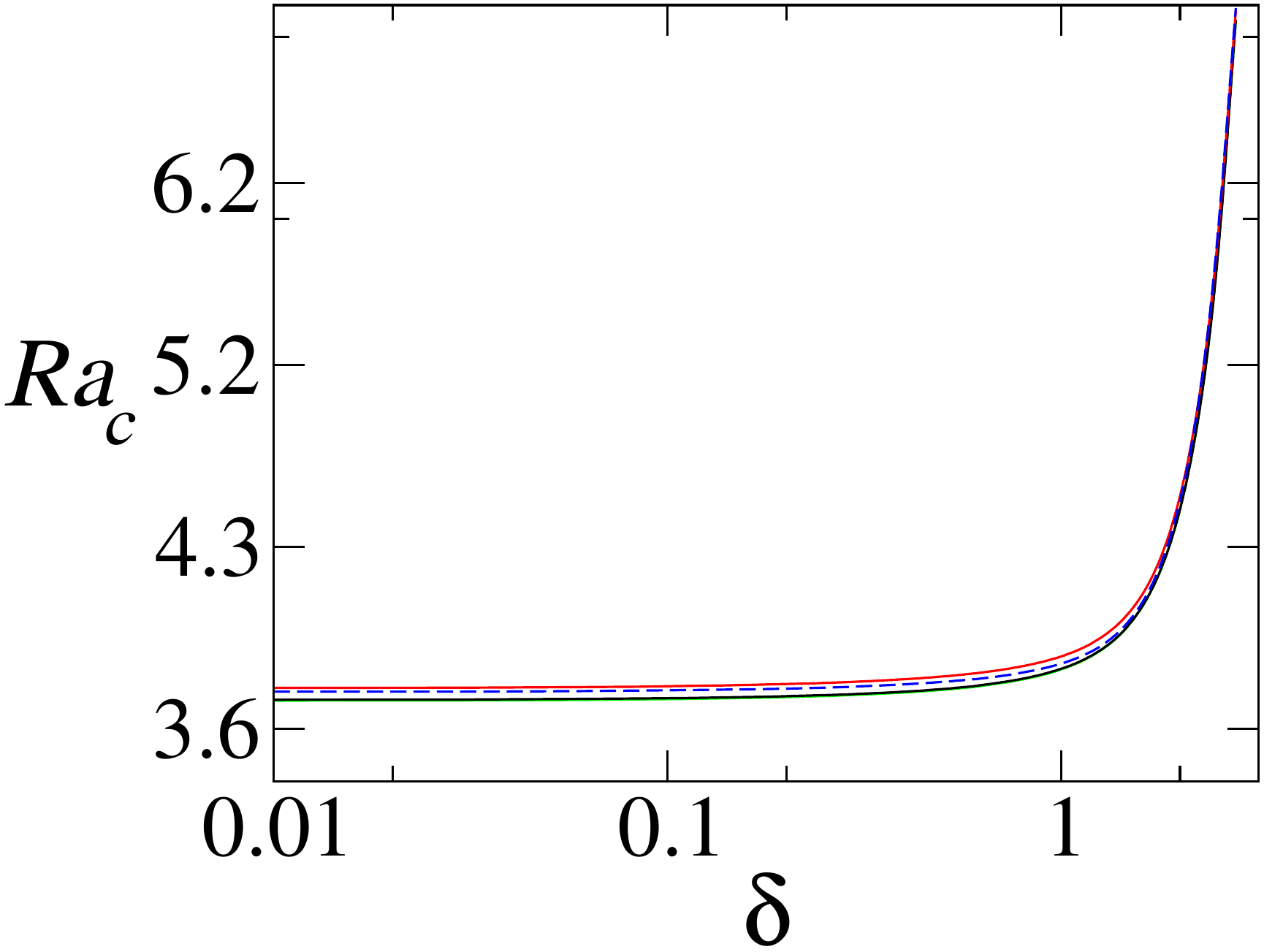}
\includegraphics[width=0.32\textwidth]{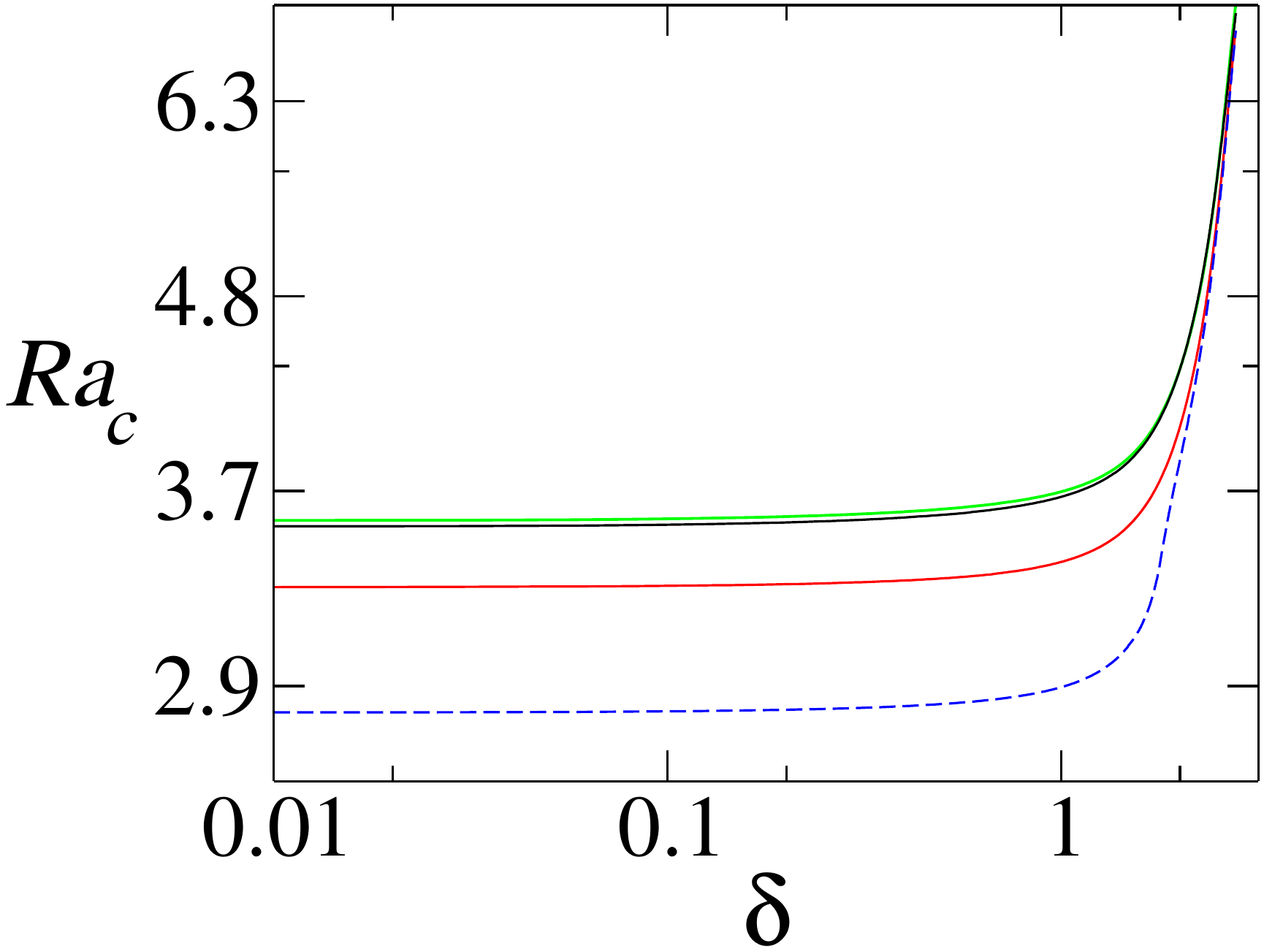}
\caption{Variation of the critical Rayleigh number $(Ra_{c})$ with $(\delta)$ for different values of the Schmidt number $(Sc)$. (a) $N=-0.5$, (b) $N=0$ and (c) $N=0.5$. The values of the remaining parameters are $Re=100$ and $Pr=0.7$.}
\label{fig6}
\end{figure*}  

\begin{figure}[h] 
\centering
\includegraphics[scale=0.45]{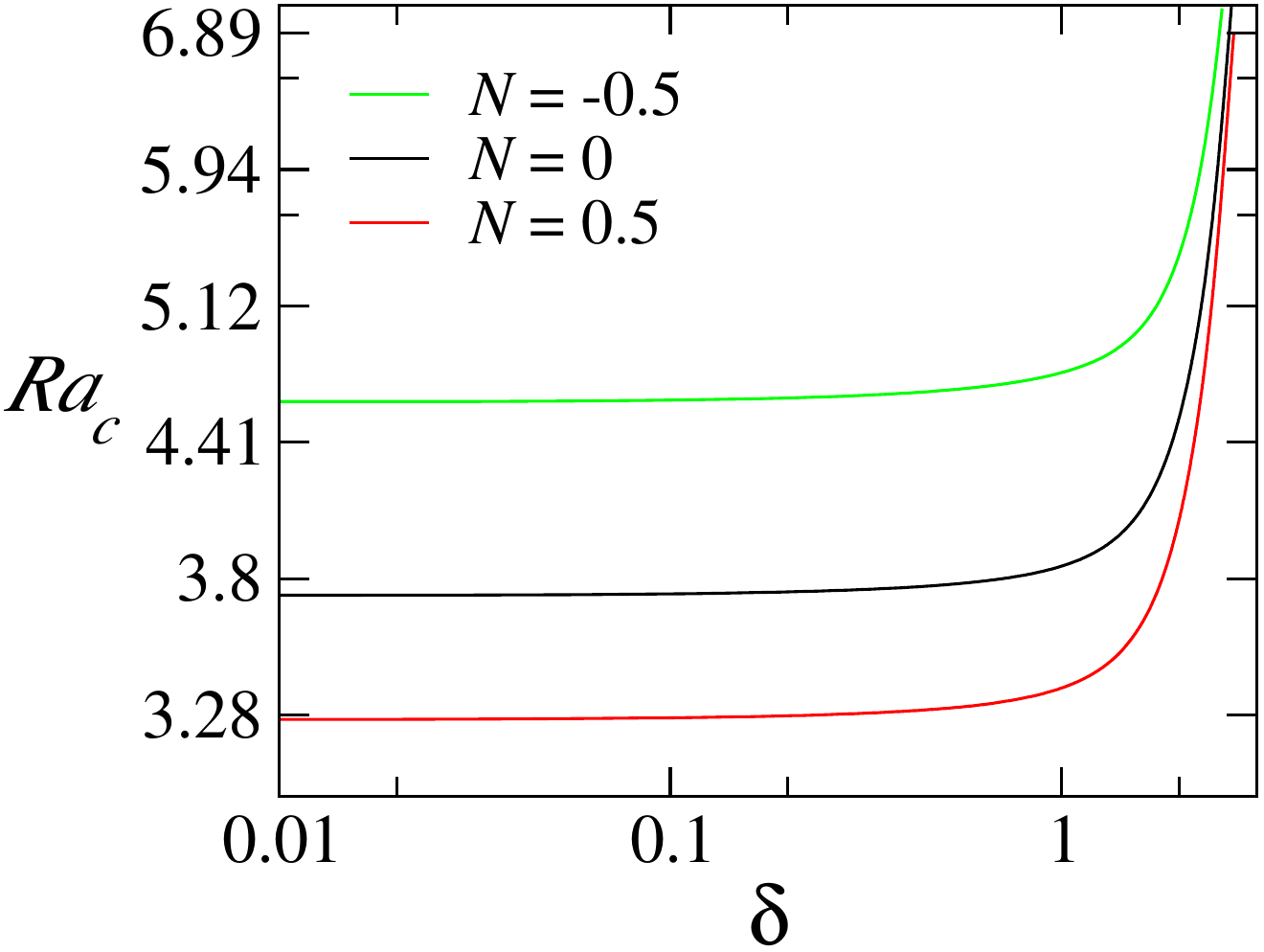}
\caption{Variation of critical Rayleigh number with $\delta$ for different values of $N$. The values of the rest of the dimensionless parameters are $Sc=1, Pr=0.7$ and $Re=100$.}
\label{fig7}
\end{figure} 

\begin{figure*} 
\centering
\hspace{0.7cm} (a) \hspace{5.2cm} (b) \hspace{5.2cm} (c)  \\
\includegraphics[width=0.33\textwidth]{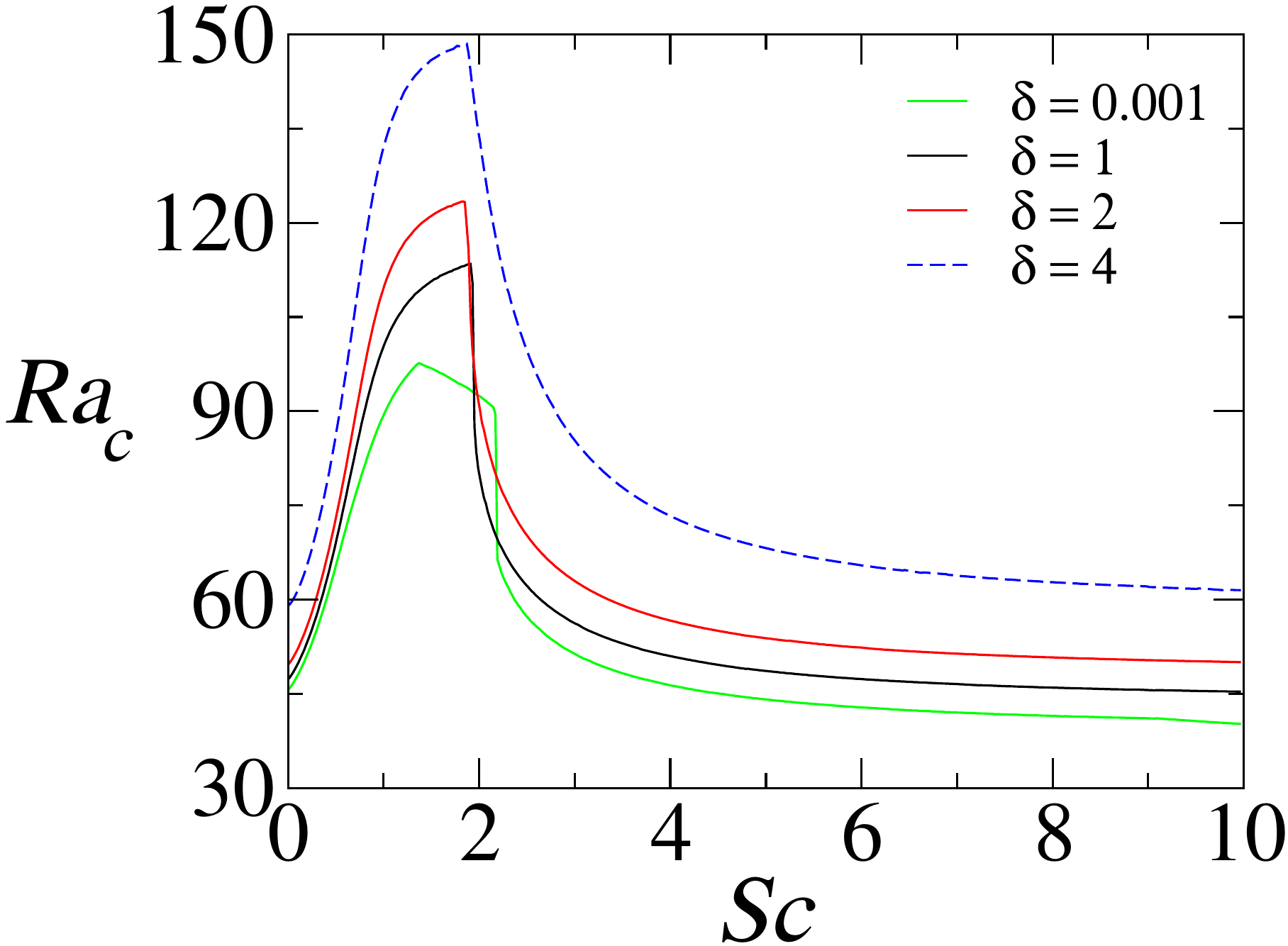}
\includegraphics[width=0.32\textwidth]{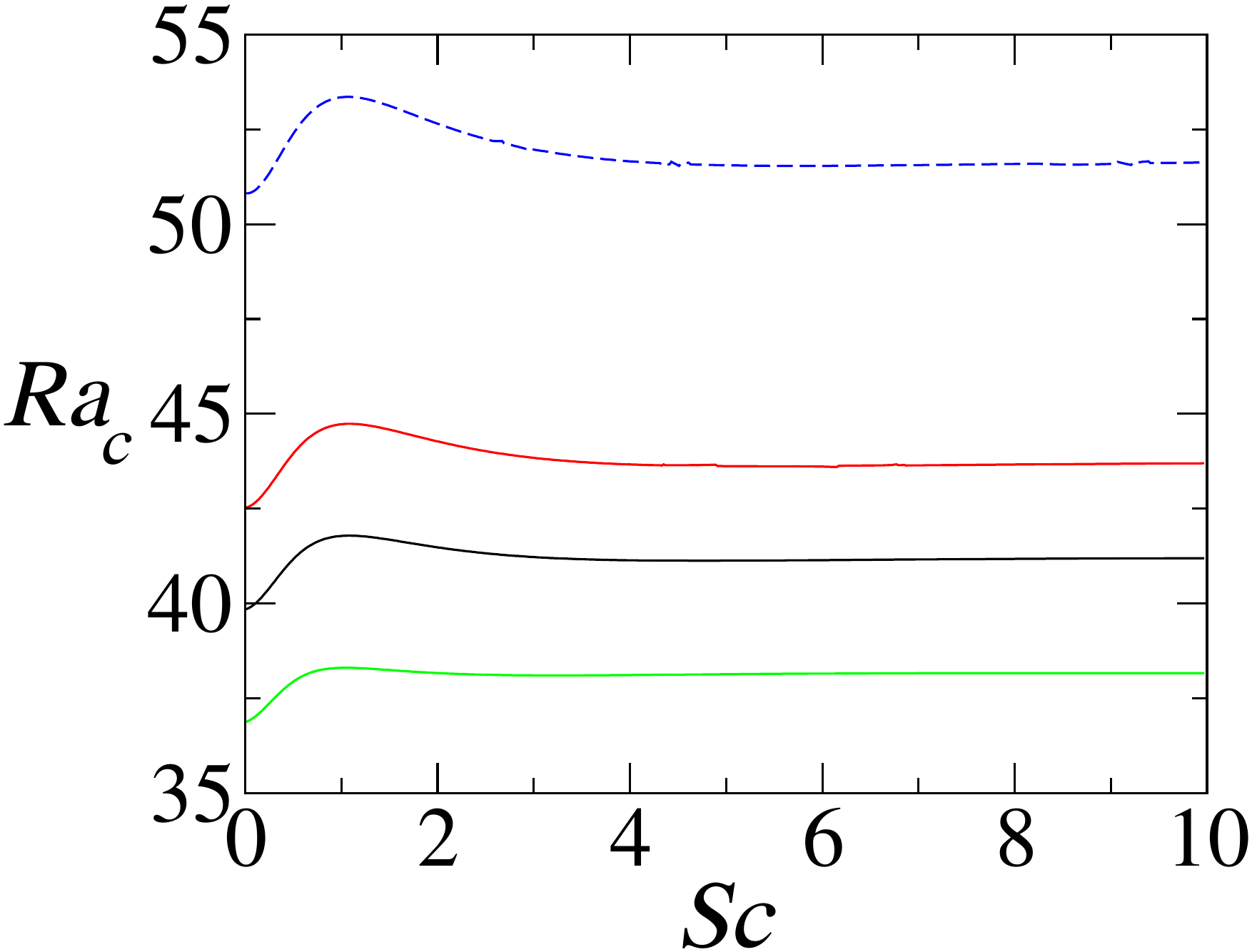}
\includegraphics[width=0.32\textwidth]{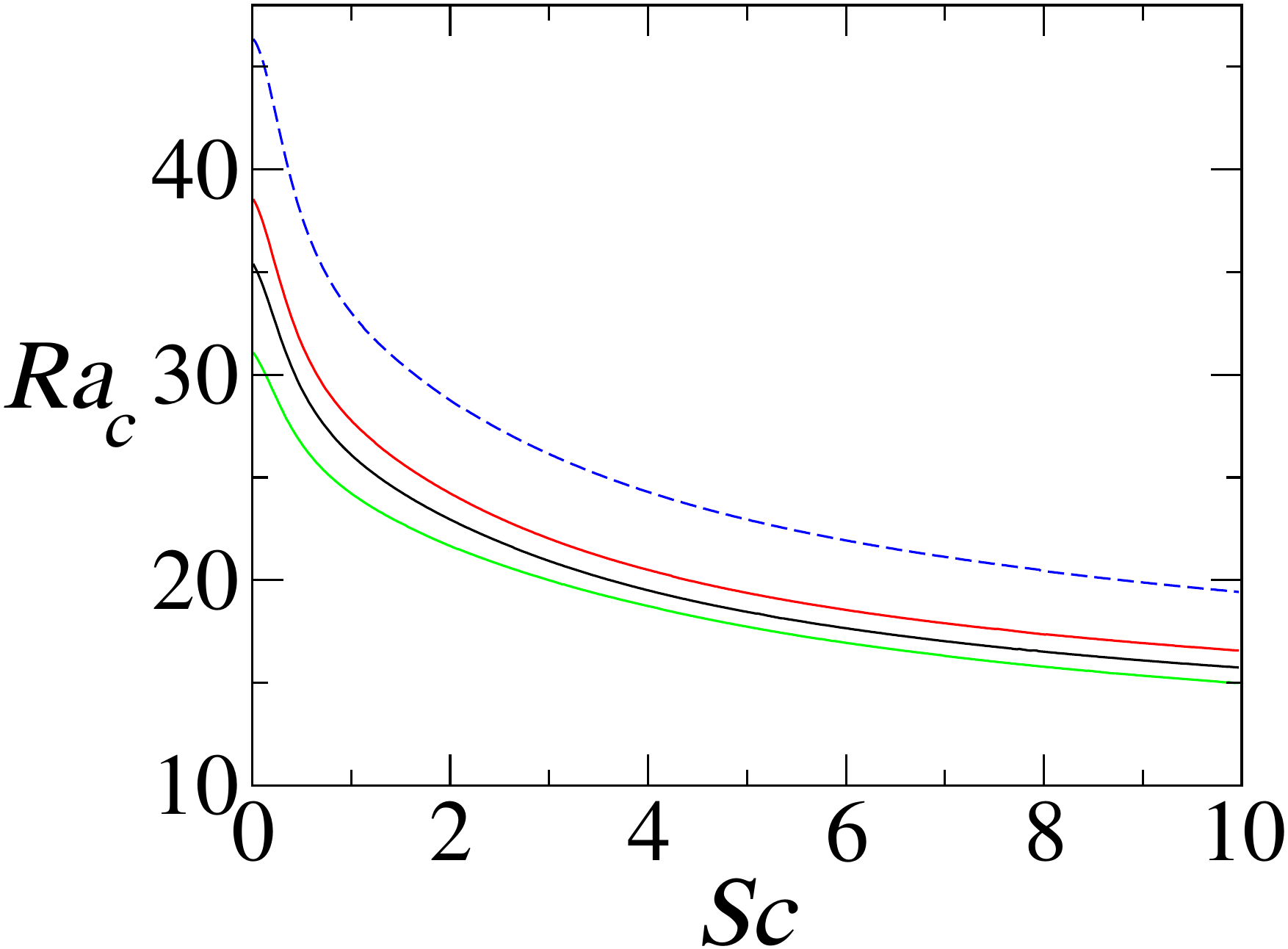}
\caption{Variation of critical Rayleigh number with $Sc$ for different values of $\delta$. (a) $N=-0.5$, (b) $N=0$ and (c) $N=0.5$. The values of the remaining parameters are $Re=100$ and $Pr=0.7$.}
\label{fig8}
\end{figure*}  

\begin{figure*} 
\centering
\hspace{0.7cm} (a) \hspace{5.2cm} (b) \hspace{5.2cm} (c)  \\
\includegraphics[width=0.32\textwidth]{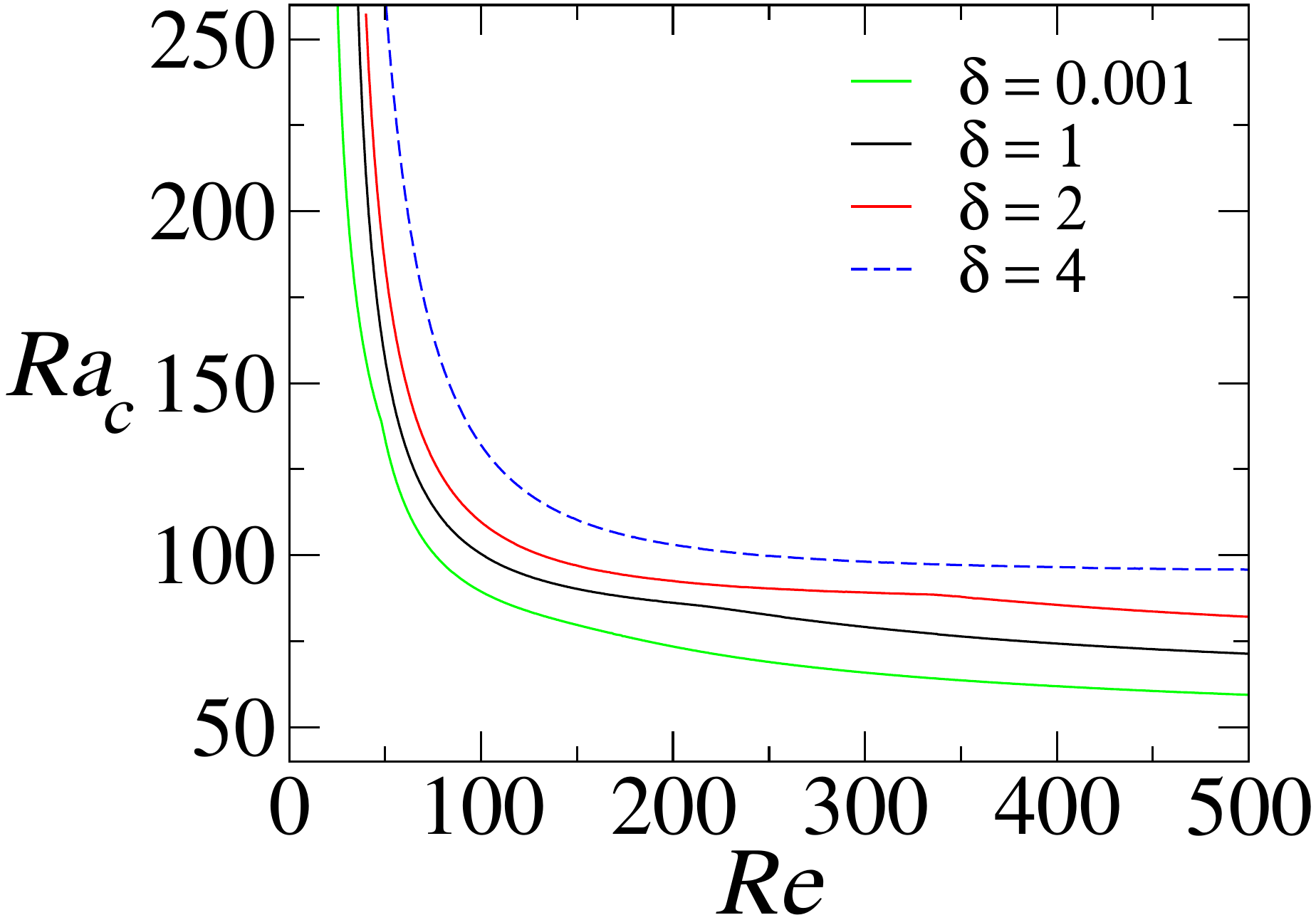}
\includegraphics[width=0.32\textwidth]{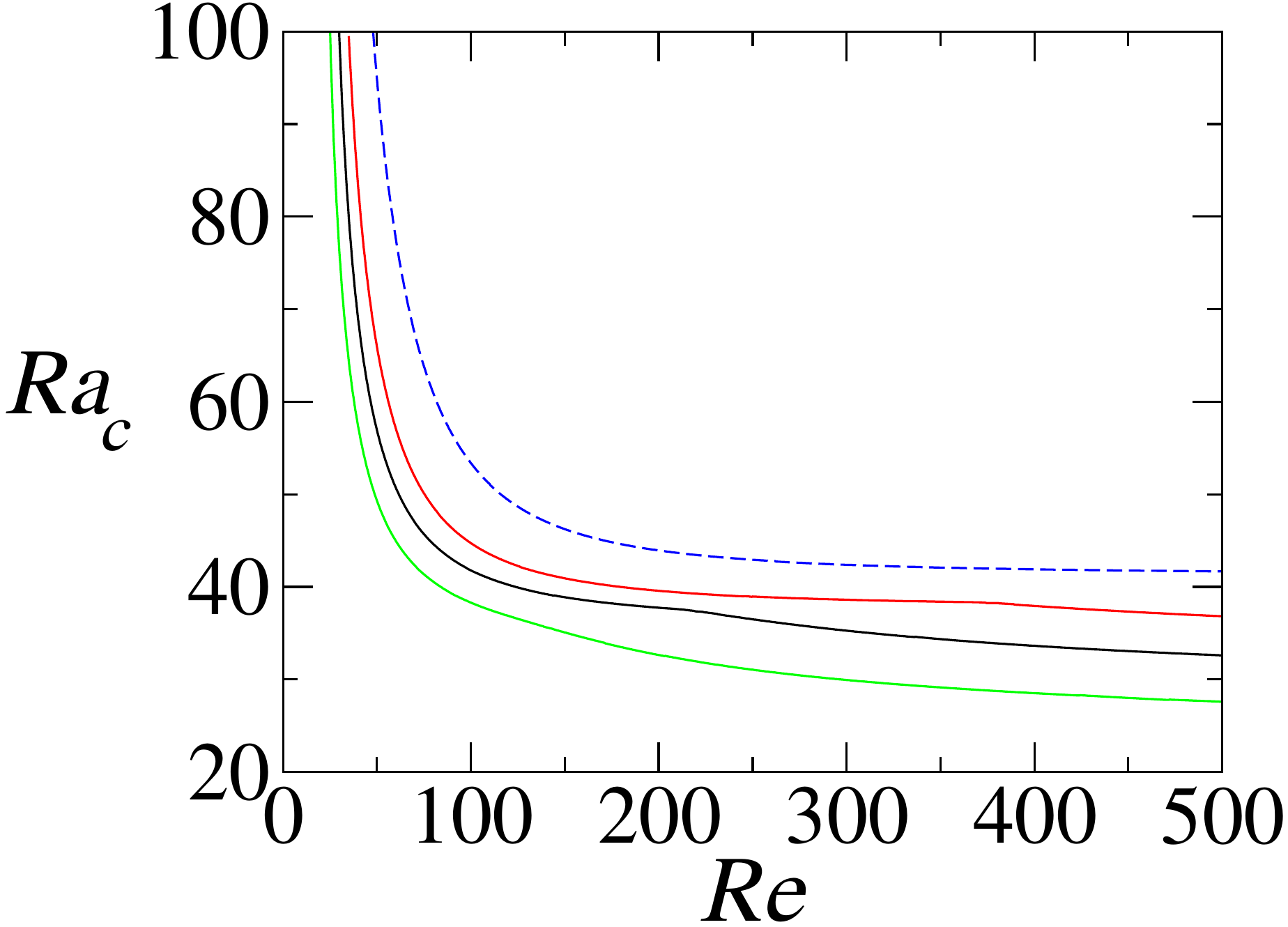}
\includegraphics[width=0.32\textwidth]{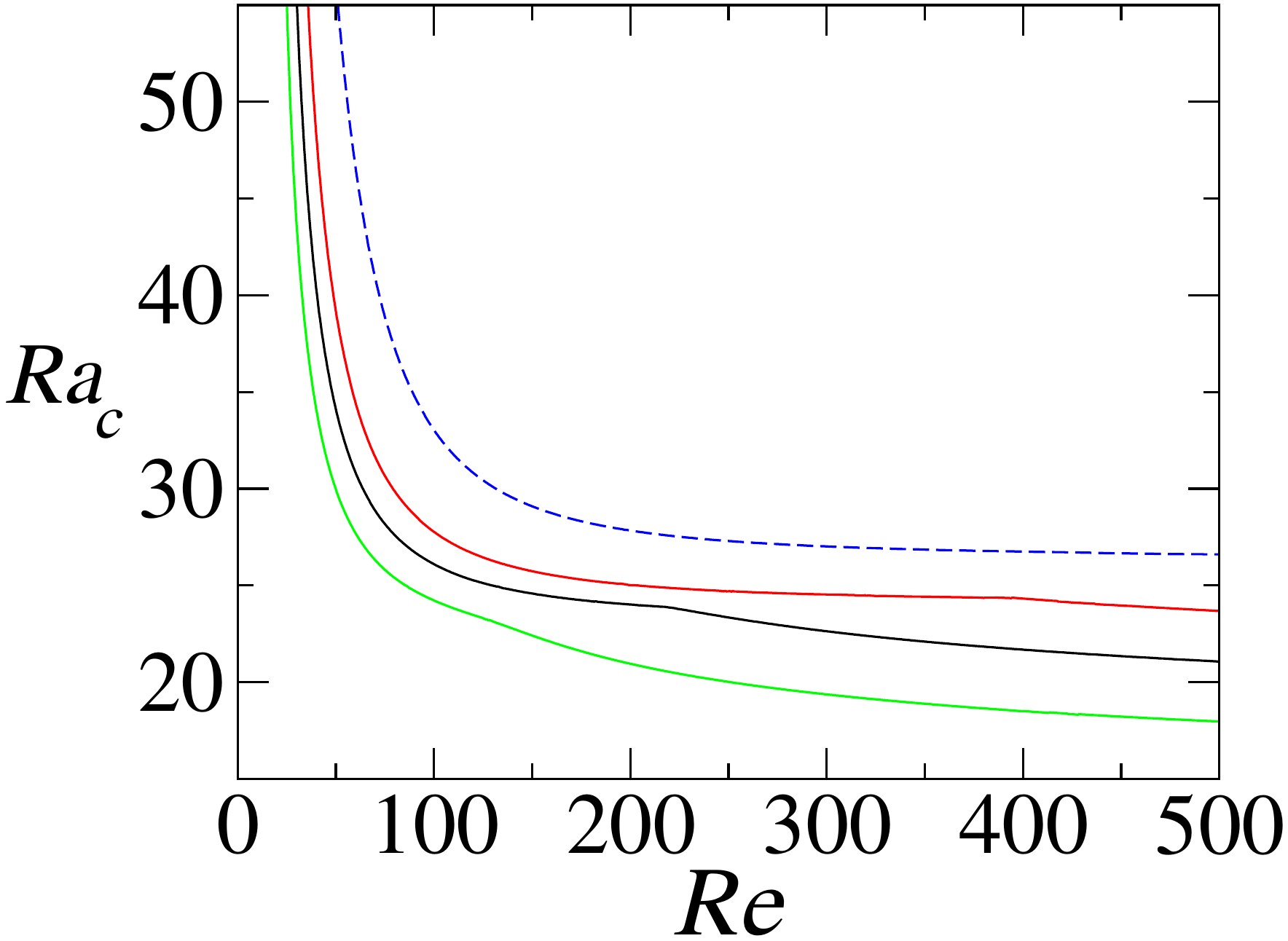}
\caption{Variation of critical Rayleigh number with $Re$ for different values of $\delta$. (a) $N=-0.5$, (b) $N=0$ and (c) $N=0.5$. The values of the remaining parameters are $Sc=1$ and $Pr=0.7$.}
\label{fig9}
\end{figure*}  

To demarcate the unstable and stable regions in $Ra-\alpha$ space, we plot the neutral stability curves (counters of $c_i=0$) for different values of the activation energy parameter ($\delta$) in the three different configurations, namely with $N=-0.5$ (figure \ref{fig5}a), $N=0$ (figure \ref{fig5}b) and $N=0.5$ (figure \ref{fig5}c). The rest of the dimensionless parameters are $Re=100$, $Pr=0.7$ and $Sc=1$. The regions below and above these curves represent the stable and unstable zones, respectively, with a $c_i=0$ boundary separating them. Figure \ref{fig5} also depicts the critical Rayleigh number ($Ra_{c}$), which is associated with the lowest value of $Ra$ for which the flow becomes unstable. It can be seen that increasing the value of $\delta$ widens the stability zone and increases the critical Rayleigh number for all values of $N$ considered in our study. This also confirms the stabilizing influence of $\delta$. By comparing the neutral stability curves in figure \ref{fig5} for different values of $N$ for a particular value of $\delta$, we observe that the critical Rayleigh number is lowest for $N=0.5$. (solutal-buoyancy-assisted flow). It shows that solutal-buoyancy-assisted flow is the least stable flow for a given set of parameters compared to solutal-buoyancy-opposed flow and the pure thermal diffusion scenario. Inspection of figure \ref{fig5} also reveals that the solutal-buoyancy-opposed flow has the narrowest range of unstable wavenumbers for $\delta = 4$. The findings discussed here corroborated the results presented in figure \ref{fig4}(a-c).

Further, to widen the range of our parametric study, the variations of the critical Rayleigh number ($Ra_c$) with $\delta$ for different values of the Schmidt number are depicted in figure \ref{fig6}(a)-(c) for $N=-0.5$, 0 and 0.5, respectively. It can be observed that the behavior of the critical Rayleigh number in the case of solutal-buoyancy-opposed flow with respect to $Sc$ is non-monotonic, as illustrated in figure \ref{fig6}(a). The flow is found to be in the most stable state when $Sc=1$, i.e., when the momentum and mass diffusion rates are equal. However, the critical Rayleigh numbers at $Sc = 0.01$, $0.1$ and 10 do not change significantly. For $\ln(\delta)> 1$, the curves corresponding to $Sc = 0.01$ and $Sc = 0.1$ are identical.  When $\ln(\delta)<1$, as in the case of pure thermal diffusion flow shown in figure \ref{fig6}(b), increasing the value of $Sc$ for a fixed $\delta$ results in a minor change in the critical value of the Rayleigh number as there is no solutal buoyancy force in this case. A thorough examination of figure \ref{fig6}(b) reveals that all curves coincide for large $\delta$ values, indicating the supremacy of viscous forces.  As seen in figure \ref{fig6}(c), for solutal-buoyancy-assisted flow, the value of the critical Rayleigh number decreases as the value of $Sc$ increases. It demonstrates how $Sc$ destabilizes the solutal-buoyancy-assisted base state flow for a particular set of parameters. It can be seen that all curves in figure \ref{fig6}(c) coincide for high values of $\delta$. It demonstrates that viscous effects dominate mass diffusivity at high $\delta$ values owing to a delay in instability. Figure \ref{fig6}(a) illustrates how $\delta$ stabilizes the base-state flow for each investigated value of $Sc$ in the case of solutal-buoyancy-opposed flow. Figure \ref{fig6}(b) and (c) also depict that pure thermal diffusion and solutal-buoyancy-assisted flow, respectively, show a similar tendency.


Figure \ref{fig7} presents the variations of critical Rayleigh number, $Ra_{c}$ with $\delta$ for different values of $N$. As seen in figure \ref{fig7}, the value of the critical Rayleigh number ($Ra_c$) is decreasing with an increase in the value of the buoyancy ratio ($N$) for a fixed value of $\delta$. Thus, the solutal-buoyancy-opposed and solutal-buoyancy-assisted base flows are associated with the most and least stable flow configurations. For each investigated value of $N$ (i.e., -0.5, 0, 0.5), we observed a slight shift in the critical value of the Rayleigh number when the range of $\ln(\delta)<1$. This indicates that in this $\delta$ range, viscosity has not significantly affected flow instability. However, a significant increase in the critical Rayleigh number is seen when $\ln(\delta)>1$. This dramatic change is caused by viscous force being more dominating when $\ln(\delta)>1$. All the curves move towards one another, which suggests that the viscous force outweighs the buoyant force, according to a detailed examination for high $\delta$ values.

The following discusses the impact of momentum diffusion versus mass diffusion on the linear instability characteristics in the viscosity-stratified flow. Under solutal-buoyancy-opposed, pure thermal diffusion, and solutal-buoyancy-assisted conditions, figure \ref{fig8}(a-c) illustrates the fluctuation of the critical Rayleigh number ($Ra_c$) with the Schmidt number $(Sc)$ for various values of the activation energy parameter ($\delta$) for $Re=100$ and $Pr=0.7$. Each curve in figure \ref{fig8}(a) illustrates how, for solutal-buoyancy-opposed flow, the critical Rayleigh number suddenly increases within a given range of $Sc$ and abruptly decreases outside that range. As a result, $Sc$ has a stabilizing effect inside that range and a destabilizing effect outside of it. The outcomes in figure \ref{fig8}(a) align with those in figure \ref{fig6}(a). In figure \ref{fig8}(b) ($N=0$), it can be seen that for all the values of $\delta$ taken into consideration, the value of $Ra_c$ rises for a small range of $Sc$ and then steadily declines until there is minimal variance. Under a pure thermal diffusion state, $Sc$ has both stabilizing and destabilizing effects within its range. Furthermore, it can be seen that $\delta$ has a stabilizing effect for $N=0$. The results for solutal-buoyancy-assisted flow are shown in figure \ref{fig8}(c). In figure \ref{fig8}(c), $Ra_c$ decreases as the value of $Sc$ increases, reflecting destabilizing behavior of the Schmidt number. This is true for every examined value of the $\delta$. \citet{khandelwal2021instabilities} discovered similar effects of the Schmidt number at the onset of instability in a solutal-buoyancy-assisted flow. In contrast, $\delta$ stabilizes, consistent with the outcomes shown in figure \ref{fig6}(c). The critical Rayleigh number decreases with an increase in the value of $N$ across the $Sc$ range. It is the smallest for solutal-buoyancy-assisted flow, making it the most unstable flow, according to the examination of the results shown in figure \ref{fig8}(a)-(c) for $\delta = 4$. Similar flow patterns have been found for other values of $\delta$.  

Finally, we present the critical Rayleigh number $(Ra_c)$ with the Reynolds number $(Re)$ for different values of the activation energy parameter ($\delta$) in figure \ref{fig9}(a)-(c) for $N=-0.5$, 0 and 0.5, respectively. In figure \ref{fig9}(a), it is clear that the value of $Ra_c$ is high at low Reynolds numbers. As we increase the value of $Re$, we observe a sharp decline in the value of $Ra_c$, followed by a slow decline after reaching a certain level of $Re$. As a result, increasing $Re$ causes the viscosity-stratified mixed-convective base-state flow to become unstable.  All investigated values of $\delta$ follow the same trend in figure \ref{fig9}(a) and (c). It also shows that the activation energy parameter $\delta$ stabilizes the whole range of $Re$. Finally, we examine how the buoyancy ratio ($N$) affects the critical Rayleigh number over the whole $Re$ range. This is accomplished by contrasting the curves in figure \ref{fig9}(a)–(c) for $\delta=0.001$, and we discovered that the value of the critical Rayleigh number decreases as the value of $N$ increases, i.e., $Ra_c$ is lowest for solutal-buoyancy-assisted flow. The same pattern holds for other considered $\delta$ values. The solutal-buoyancy-assisted flow is shown to be the most unstable flow.

\section{Concluding remarks}
\label{sec:conc}
We have conducted a numerical investigation to analyze the linear stability of a pressure-driven viscosity stratified flow in a vertical channel under the influence of double-diffusive mixed convection. The viscosity, which is a function of temperature and concentration, is defined using \ks{the Nahme-type viscosity-temperature relationship}. The Chebyshev spectral collocation method is used to solve the eigenvalue equations derived from linear stability analyses. We examine the impact of varying the activation energy parameter ($\delta$), Reynolds number ($Re$), and Schmidt number ($Sc$) on the linear stability characteristics in three different scenarios: (i) solutal-buoyancy-opposed flow ($N < 0$), (ii) flow resulting solely from thermal diffusion ($N = 0$), and (iii) solutal-buoyancy-assisted flow ($N > 0$). The growth rate profiles for $N=-0.5$, 0 and 0.5 reveal that increasing the activation energy parameter ($\delta$) results in a reduction in the maximum growth rate of the disturbances, indicating the stabilizing effect of $\delta$. Additionally, positive growth rates in a certain range of wavenumbers indicate the unstable behavior of the base state flow for the considered parameters. Increasing $\delta$ for all values of $N$ delays the onset of convection. In the cases of $N=-0.5$ and 0, both stabilizing and destabilizing behavior of $Sc$ is observed, while only destabilizing behavior of $Sc$ is observed for $N=0.5$. Moreover, the buoyancy-assisted flow is the most unstable flow, and increasing the Reynolds number for all values of $\delta$ and $N$ reduces the stability of the flow as expected.

 \section*{Acknowledgements:} K.C.S. thanks the Science \& Engineering Research Board and IIT Hyderabad, India for their financial supports through grants CRG/2020/000507 and IITH/CHE/F011/SOCH1, respectively. Ankush acknowledges his gratitude to University Grants Commission (UGC), India, for providing financial assistance.


\section*{Data Availability Statement}
The data that support the findings of this study are available from the corresponding author upon reasonable request.

\clearpage
\ks{
\appendix*
\section{Derivation of the dimensionless energy equation}
The energy equation in the dimensional form is given by 
\begin{equation}
     {\partial T \over \partial t}+\textbf{V}.\nabla T= D \nabla^2T, \label{eq21}
\end{equation}
For convenience, first, we non-dimensionalize the left hand side (L.H.S.) of equation (\ref{eq21}) using equation. (\ref{eq2}).
\begin{eqnarray}
    L.H.S. = {\partial (\theta A_{1}HPrRe + T_w) \over \partial \widetilde{t}}{\partial \widetilde{t} \over \partial t} + \nonumber \\ \overline{U}\widetilde{u}{\partial (\theta A_{1}HPrRe + T_{w}) \over \partial \widetilde{x}}{\partial \widetilde{x} \over \partial x} + \nonumber \\ \overline{U}\widetilde{v}{\partial (\theta A_{1}HPrRe + T_{w}) \over \partial \widetilde{y}}{\partial \widetilde{y} \over \partial y} + \nonumber \\ \overline{U}\widetilde{w}{\partial (\theta A_{1}HPrRe + T_{w}) \over \partial \widetilde{z}}{\partial \widetilde{z} \over \partial z}.\\
\end{eqnarray}
Substituting the expression $T_{w} = T_0 + A_1 x$, we get
\begin{eqnarray}
     L.H.S.={\partial (\theta A_{1}HPrRe + T_0 + A_{1}\widetilde{x}H) \over \partial \widetilde{t}}{\partial \widetilde{t} \over \partial t} + \nonumber \\ \overline{U}\widetilde{u}{\partial (\theta A_{1}HPrRe + T_0 + A_{1}\widetilde{x}H) \over \partial \widetilde{x}}{\partial \widetilde{x} \over \partial x} \nonumber \\ + \overline{U}\widetilde{y}{\partial (\theta A_{1}HPrRe + T_0 + A_{1}\widetilde{x}H) \over \partial \widetilde{y}}{\partial \widetilde{y} \over \partial y} + \nonumber \\ \overline{U}\widetilde{w}{\partial (\theta A_{1}HPrRe + T_0 + A_{1}\widetilde{x}H) \over \partial \widetilde{z}}{\partial \widetilde{z} \over \partial z}.
\end{eqnarray}
By rearranging, we get
\begin{eqnarray}
L.H.S.=\overline{U}A_{1}PrRe{\partial \theta \over \partial \widetilde{t}} + \overline{U}A_{1}PrRe\widetilde{u}{\partial \theta \over \partial \widetilde{x}} + \overline{U}A_{1}\widetilde{u} +  \nonumber \\ \overline{U}A_{1}PrRe\widetilde{v}{\partial \theta \over \partial \widetilde{y}} +  \overline{U}A_{1}PrRe\widetilde{w}{\partial \theta \over \partial \widetilde{z}}.~~~~~~~~~~~~~~~~~
\end{eqnarray}
In vector form, this equation can be written as 
\begin{equation}
    L.H.S.=\overline{U}A_{1}PrRe{\partial \theta \over \partial \widetilde{t}} + \overline{U}A_{1}PrRe\widetilde{\textbf{V}}.\nabla \theta + \overline{U}A_{1}\widetilde{u}. \label{eq26}
\end{equation}
Then, following the same procedure for the non-dimensionalisation of the right hand side (R.H.S.) of equation (\ref{eq21}) using equation (\ref{eq2}), we get
\begin{equation}
    R.H.S.={D A_{1}PrRe \over H}\nabla^2\theta. \label{eq27}
\end{equation}
Combining equations (\ref{eq26}) and (\ref{eq27}), we get the dimensionless form of the energy equation as
\begin{equation}
{\partial \theta \over \partial \widetilde{t}} + \widetilde{\textbf{V}}.\nabla \theta ={1 \over RePr}(\nabla^2\theta - \widetilde{u}).
\end{equation}
In the Cartesian index notation, this equation is expressed as equation (\ref{eq5}). By following a similar derivation, we can also obtain the dimensionless concentration-diffusion equation (\ref{eq6}).
}

\section*{REFERENCES}

\begin{thebibliography}{49}%
\makeatletter
\providecommand \@ifxundefined [1]{%
 \@ifx{#1\undefined}
}%
\providecommand \@ifnum [1]{%
 \ifnum #1\expandafter \@firstoftwo
 \else \expandafter \@secondoftwo
 \fi
}%
\providecommand \@ifx [1]{%
 \ifx #1\expandafter \@firstoftwo
 \else \expandafter \@secondoftwo
 \fi
}%
\providecommand \natexlab [1]{#1}%
\providecommand \enquote  [1]{``#1''}%
\providecommand \bibnamefont  [1]{#1}%
\providecommand \bibfnamefont [1]{#1}%
\providecommand \citenamefont [1]{#1}%
\providecommand \href@noop [0]{\@secondoftwo}%
\providecommand \href [0]{\begingroup \@sanitize@url \@href}%
\providecommand \@href[1]{\@@startlink{#1}\@@href}%
\providecommand \@@href[1]{\endgroup#1\@@endlink}%
\providecommand \@sanitize@url [0]{\catcode `\\12\catcode `\$12\catcode
  `\&12\catcode `\#12\catcode `\^12\catcode `\_12\catcode `\%12\relax}%
\providecommand \@@startlink[1]{}%
\providecommand \@@endlink[0]{}%
\providecommand \url  [0]{\begingroup\@sanitize@url \@url }%
\providecommand \@url [1]{\endgroup\@href {#1}{\urlprefix }}%
\providecommand \urlprefix  [0]{URL }%
\providecommand \Eprint [0]{\href }%
\providecommand \doibase [0]{http://dx.doi.org/}%
\providecommand \selectlanguage [0]{\@gobble}%
\providecommand \bibinfo  [0]{\@secondoftwo}%
\providecommand \bibfield  [0]{\@secondoftwo}%
\providecommand \translation [1]{[#1]}%
\providecommand \BibitemOpen [0]{}%
\providecommand \bibitemStop [0]{}%
\providecommand \bibitemNoStop [0]{.\EOS\space}%
\providecommand \EOS [0]{\spacefactor3000\relax}%
\providecommand \BibitemShut  [1]{\csname bibitem#1\endcsname}%
\let\auto@bib@innerbib\@empty
\bibitem [{\citenamefont {Joseph}\ \emph {et~al.}(1997)\citenamefont {Joseph},
  \citenamefont {Bai}, \citenamefont {Chen},\ and\ \citenamefont
  {Renardy}}]{joseph1997core}%
  \BibitemOpen
  \bibfield  {author} {\bibinfo {author} {\bibfnamefont {D.~D.}\ \bibnamefont
  {Joseph}}, \bibinfo {author} {\bibfnamefont {R.}~\bibnamefont {Bai}},
  \bibinfo {author} {\bibfnamefont {K.~P.}\ \bibnamefont {Chen}}, \ and\
  \bibinfo {author} {\bibfnamefont {Y.~Y.}\ \bibnamefont {Renardy}},\
  }\bibfield  {title} {\enquote {\bibinfo {title} {Core-annular flows},}\
  }\href@noop {} {\bibfield  {journal} {\bibinfo  {journal} {Annual Review of
  Fluid Mechanics}\ }\textbf {\bibinfo {volume} {29}},\ \bibinfo {pages}
  {65--90} (\bibinfo {year} {1997})}\BibitemShut {NoStop}%
\bibitem [{\citenamefont {Selvam}\ \emph {et~al.}(2007)\citenamefont {Selvam},
  \citenamefont {Merk}, \citenamefont {Govindarajan},\ and\ \citenamefont
  {Meiburg}}]{selvam2007stability}%
  \BibitemOpen
  \bibfield  {author} {\bibinfo {author} {\bibfnamefont {B.}~\bibnamefont
  {Selvam}}, \bibinfo {author} {\bibfnamefont {S.}~\bibnamefont {Merk}},
  \bibinfo {author} {\bibfnamefont {R.}~\bibnamefont {Govindarajan}}, \ and\
  \bibinfo {author} {\bibfnamefont {E.}~\bibnamefont {Meiburg}},\ }\bibfield
  {title} {\enquote {\bibinfo {title} {Stability of miscible core--annular
  flows with viscosity stratification},}\ }\href@noop {} {\bibfield  {journal}
  {\bibinfo  {journal} {J. Fluid Mech.}\ }\textbf {\bibinfo {volume} {592}},\
  \bibinfo {pages} {23--49} (\bibinfo {year} {2007})}\BibitemShut {NoStop}%
\bibitem [{\citenamefont {Govindarajan}\ and\ \citenamefont
  {Sahu}(2014)}]{govindarajan2014instabilities}%
  \BibitemOpen
  \bibfield  {author} {\bibinfo {author} {\bibfnamefont {R.}~\bibnamefont
  {Govindarajan}}\ and\ \bibinfo {author} {\bibfnamefont {K.~C.}\ \bibnamefont
  {Sahu}},\ }\bibfield  {title} {\enquote {\bibinfo {title} {Instabilities in
  viscosity-stratified flow},}\ }\href@noop {} {\bibfield  {journal} {\bibinfo
  {journal} {Annu. Rev. Fluid Mech.}\ }\textbf {\bibinfo {volume} {46}},\
  \bibinfo {pages} {331--353} (\bibinfo {year} {2014})}\BibitemShut {NoStop}%
\bibitem [{\citenamefont {Chen}\ and\ \citenamefont
  {Meiburg}(1996)}]{chen1996miscible}%
  \BibitemOpen
  \bibfield  {author} {\bibinfo {author} {\bibfnamefont {C.~Y.}\ \bibnamefont
  {Chen}}\ and\ \bibinfo {author} {\bibfnamefont {E.}~\bibnamefont {Meiburg}},\
  }\bibfield  {title} {\enquote {\bibinfo {title} {Miscible displacements in
  capillary tubes. {P}art 2. numerical simulations},}\ }\href@noop {}
  {\bibfield  {journal} {\bibinfo  {journal} {J. Fluid Mech.}\ }\textbf
  {\bibinfo {volume} {326}},\ \bibinfo {pages} {57--90} (\bibinfo {year}
  {1996})}\BibitemShut {NoStop}%
\bibitem [{\citenamefont {Petitjeans}\ and\ \citenamefont
  {Maxworthy}(1996)}]{petitjeans1996miscible}%
  \BibitemOpen
  \bibfield  {author} {\bibinfo {author} {\bibfnamefont {P.}~\bibnamefont
  {Petitjeans}}\ and\ \bibinfo {author} {\bibfnamefont {T.}~\bibnamefont
  {Maxworthy}},\ }\bibfield  {title} {\enquote {\bibinfo {title} {Miscible
  displacements in capillary tubes. {P}art 1. experiments},}\ }\href@noop {}
  {\bibfield  {journal} {\bibinfo  {journal} {J. Fluid Mech.}\ }\textbf
  {\bibinfo {volume} {326}},\ \bibinfo {pages} {37--56} (\bibinfo {year}
  {1996})}\BibitemShut {NoStop}%
\bibitem [{\citenamefont {Sahu}\ \emph {et~al.}(2009)\citenamefont {Sahu},
  \citenamefont {Ding}, \citenamefont {Valluri},\ and\ \citenamefont
  {Matar}}]{sahu2009linear}%
  \BibitemOpen
  \bibfield  {author} {\bibinfo {author} {\bibfnamefont {K.~C.}\ \bibnamefont
  {Sahu}}, \bibinfo {author} {\bibfnamefont {H.}~\bibnamefont {Ding}}, \bibinfo
  {author} {\bibfnamefont {P.}~\bibnamefont {Valluri}}, \ and\ \bibinfo
  {author} {\bibfnamefont {O.~K.}\ \bibnamefont {Matar}},\ }\bibfield  {title}
  {\enquote {\bibinfo {title} {Linear stability analysis and numerical
  simulation of miscible two-layer channel flow},}\ }\href@noop {} {\bibfield
  {journal} {\bibinfo  {journal} {Phys. Fluids}\ }\textbf {\bibinfo {volume}
  {21}},\ \bibinfo {pages} {042104} (\bibinfo {year} {2009})}\BibitemShut
  {NoStop}%
\bibitem [{\citenamefont {Pearson}(1985)}]{pearson1985mechanics}%
  \BibitemOpen
  \bibfield  {author} {\bibinfo {author} {\bibfnamefont {J.~R.}\ \bibnamefont
  {Pearson}},\ }\href@noop {} {\emph {\bibinfo {title} {Mechanics of polymer
  processing}}}\ (\bibinfo  {publisher} {Springer Science \& Business Media},\
  \bibinfo {year} {1985})\BibitemShut {NoStop}%
\bibitem [{\citenamefont {Cao}\ \emph {et~al.}(2003)\citenamefont {Cao},
  \citenamefont {Ventresca}, \citenamefont {Sreenivas},\ and\ \citenamefont
  {Prasad}}]{cao2003instability}%
  \BibitemOpen
  \bibfield  {author} {\bibinfo {author} {\bibfnamefont {Q.}~\bibnamefont
  {Cao}}, \bibinfo {author} {\bibfnamefont {A.~L.}\ \bibnamefont {Ventresca}},
  \bibinfo {author} {\bibfnamefont {K.~R.}\ \bibnamefont {Sreenivas}}, \ and\
  \bibinfo {author} {\bibfnamefont {A.~K.}\ \bibnamefont {Prasad}},\ }\bibfield
   {title} {\enquote {\bibinfo {title} {Instability due to viscosity
  stratification downstream of a centerline injector},}\ }\href@noop {}
  {\bibfield  {journal} {\bibinfo  {journal} {Can. J. Chem. Eng.}\ }\textbf
  {\bibinfo {volume} {81}},\ \bibinfo {pages} {913--922} (\bibinfo {year}
  {2003})}\BibitemShut {NoStop}%
\bibitem [{\citenamefont {Regner}\ \emph {et~al.}(2007)\citenamefont {Regner},
  \citenamefont {Henningsson}, \citenamefont {Wiklund}, \citenamefont
  {{\"O}stergren},\ and\ \citenamefont
  {Tr{\"a}g{\aa}rdh}}]{regner2007predicting}%
  \BibitemOpen
  \bibfield  {author} {\bibinfo {author} {\bibfnamefont {M.}~\bibnamefont
  {Regner}}, \bibinfo {author} {\bibfnamefont {M.}~\bibnamefont {Henningsson}},
  \bibinfo {author} {\bibfnamefont {J.}~\bibnamefont {Wiklund}}, \bibinfo
  {author} {\bibfnamefont {K.}~\bibnamefont {{\"O}stergren}}, \ and\ \bibinfo
  {author} {\bibfnamefont {C.}~\bibnamefont {Tr{\"a}g{\aa}rdh}},\ }\bibfield
  {title} {\enquote {\bibinfo {title} {Predicting the displacement of yoghurt
  by water in a pipe using {CFD}},}\ }\href@noop {} {\bibfield  {journal}
  {\bibinfo  {journal} {Chem. Eng. Technol.}\ }\textbf {\bibinfo {volume}
  {30}},\ \bibinfo {pages} {844--853} (\bibinfo {year} {2007})}\BibitemShut
  {NoStop}%
\bibitem [{\citenamefont {Williams}\ \emph {et~al.}(2020)\citenamefont
  {Williams}, \citenamefont {Lee}, \citenamefont {Apriceno}, \citenamefont
  {Sear},\ and\ \citenamefont {Battaglia}}]{williams2020diffusioosmotic}%
  \BibitemOpen
  \bibfield  {author} {\bibinfo {author} {\bibfnamefont {I.}~\bibnamefont
  {Williams}}, \bibinfo {author} {\bibfnamefont {S.}~\bibnamefont {Lee}},
  \bibinfo {author} {\bibfnamefont {A.}~\bibnamefont {Apriceno}}, \bibinfo
  {author} {\bibfnamefont {R.~P.}\ \bibnamefont {Sear}}, \ and\ \bibinfo
  {author} {\bibfnamefont {G.}~\bibnamefont {Battaglia}},\ }\bibfield  {title}
  {\enquote {\bibinfo {title} {Diffusioosmotic and convective flows induced by
  a nonelectrolyte concentration gradient},}\ }\href@noop {} {\bibfield
  {journal} {\bibinfo  {journal} {Proc. Natl. Acad. Sci. U.S.A.}\ }\textbf
  {\bibinfo {volume} {117}},\ \bibinfo {pages} {25263--25271} (\bibinfo {year}
  {2020})}\BibitemShut {NoStop}%
\bibitem [{\citenamefont {Hu}\ \emph {et~al.}(2021)\citenamefont {Hu},
  \citenamefont {Huang}, \citenamefont {Zhang}, \citenamefont {Wang},\ and\
  \citenamefont {Chen}}]{hu2021nanofluid}%
  \BibitemOpen
  \bibfield  {author} {\bibinfo {author} {\bibfnamefont {K.-X.}\ \bibnamefont
  {Hu}}, \bibinfo {author} {\bibfnamefont {Y.}~\bibnamefont {Huang}}, \bibinfo
  {author} {\bibfnamefont {X.-Y.}\ \bibnamefont {Zhang}}, \bibinfo {author}
  {\bibfnamefont {S.}~\bibnamefont {Wang}}, \ and\ \bibinfo {author}
  {\bibfnamefont {Q.-S.}\ \bibnamefont {Chen}},\ }\bibfield  {title} {\enquote
  {\bibinfo {title} {The nanofluid flows in the channel with linearly varying
  wall temperature},}\ }\href@noop {} {\bibfield  {journal} {\bibinfo
  {journal} {Case Stud. Therm. Eng.}\ }\textbf {\bibinfo {volume} {28}},\
  \bibinfo {pages} {101602} (\bibinfo {year} {2021})}\BibitemShut {NoStop}%
\bibitem [{\citenamefont {Nazir}\ \emph {et~al.}(2021)\citenamefont {Nazir},
  \citenamefont {Sohail}, \citenamefont {Ali}, \citenamefont {Sherif},
  \citenamefont {Park}, \citenamefont {Lee}, \citenamefont {Selim},\ and\
  \citenamefont {Thounthong}}]{nazir2021applications}%
  \BibitemOpen
  \bibfield  {author} {\bibinfo {author} {\bibfnamefont {U.}~\bibnamefont
  {Nazir}}, \bibinfo {author} {\bibfnamefont {M.}~\bibnamefont {Sohail}},
  \bibinfo {author} {\bibfnamefont {U.}~\bibnamefont {Ali}}, \bibinfo {author}
  {\bibfnamefont {E.~M.}\ \bibnamefont {Sherif}}, \bibinfo {author}
  {\bibfnamefont {C.}~\bibnamefont {Park}}, \bibinfo {author} {\bibfnamefont
  {J.~R.}\ \bibnamefont {Lee}}, \bibinfo {author} {\bibfnamefont {M.~M.}\
  \bibnamefont {Selim}}, \ and\ \bibinfo {author} {\bibfnamefont
  {P.}~\bibnamefont {Thounthong}},\ }\bibfield  {title} {\enquote {\bibinfo
  {title} {Applications of cattaneo--christov fluxes on modelling the boundary
  value problem of prandtl fluid comprising variable properties},}\ }\href@noop
  {} {\bibfield  {journal} {\bibinfo  {journal} {Scientific Reports}\ }\textbf
  {\bibinfo {volume} {11}},\ \bibinfo {pages} {17837} (\bibinfo {year}
  {2021})}\BibitemShut {NoStop}%
\bibitem [{\citenamefont {Chen}\ and\ \citenamefont
  {Chung}(1996)}]{chen1996linear}%
  \BibitemOpen
  \bibfield  {author} {\bibinfo {author} {\bibfnamefont {Y.~C.}\ \bibnamefont
  {Chen}}\ and\ \bibinfo {author} {\bibfnamefont {J.~N.}\ \bibnamefont
  {Chung}},\ }\bibfield  {title} {\enquote {\bibinfo {title} {The linear
  stability of mixed convection in a vertical channel flow},}\ }\href@noop {}
  {\bibfield  {journal} {\bibinfo  {journal} {J. Fluid Mech.}\ }\textbf
  {\bibinfo {volume} {325}},\ \bibinfo {pages} {29--51} (\bibinfo {year}
  {1996})}\BibitemShut {NoStop}%
\bibitem [{\citenamefont {Yih}(1967)}]{yih1967instability}%
  \BibitemOpen
  \bibfield  {author} {\bibinfo {author} {\bibfnamefont {C.~S.}\ \bibnamefont
  {Yih}},\ }\bibfield  {title} {\enquote {\bibinfo {title} {Instability due to
  viscosity stratification},}\ }\href@noop {} {\bibfield  {journal} {\bibinfo
  {journal} {J. Fluid Mech.}\ }\textbf {\bibinfo {volume} {27}},\ \bibinfo
  {pages} {337--352} (\bibinfo {year} {1967})}\BibitemShut {NoStop}%
\bibitem [{\citenamefont {Mu}\ \emph {et~al.}(2021)\citenamefont {Mu},
  \citenamefont {Qiao}, \citenamefont {Si}, \citenamefont {Cheng},\ and\
  \citenamefont {Ding}}]{mu2021interfacial}%
  \BibitemOpen
  \bibfield  {author} {\bibinfo {author} {\bibfnamefont {K.}~\bibnamefont
  {Mu}}, \bibinfo {author} {\bibfnamefont {R.}~\bibnamefont {Qiao}}, \bibinfo
  {author} {\bibfnamefont {T.}~\bibnamefont {Si}}, \bibinfo {author}
  {\bibfnamefont {X.}~\bibnamefont {Cheng}}, \ and\ \bibinfo {author}
  {\bibfnamefont {H.}~\bibnamefont {Ding}},\ }\bibfield  {title} {\enquote
  {\bibinfo {title} {Interfacial instability and transition of jetting and
  dripping modes in a co-flow focusing process},}\ }\href@noop {} {\bibfield
  {journal} {\bibinfo  {journal} {Phys. Fluids}\ }\textbf {\bibinfo {volume}
  {33}},\ \bibinfo {pages} {052118} (\bibinfo {year} {2021})}\BibitemShut
  {NoStop}%
\bibitem [{\citenamefont {Potter}\ and\ \citenamefont
  {Graber}(1972)}]{potter1972stability}%
  \BibitemOpen
  \bibfield  {author} {\bibinfo {author} {\bibfnamefont {M.~C.}\ \bibnamefont
  {Potter}}\ and\ \bibinfo {author} {\bibfnamefont {E.}~\bibnamefont
  {Graber}},\ }\bibfield  {title} {\enquote {\bibinfo {title} {Stability of
  plane {P}oiseuille flow with heat transfer},}\ }\href@noop {} {\bibfield
  {journal} {\bibinfo  {journal} {Phys. Fluids}\ }\textbf {\bibinfo {volume}
  {15}},\ \bibinfo {pages} {387--391} (\bibinfo {year} {1972})}\BibitemShut
  {NoStop}%
\bibitem [{\citenamefont {Pinarbasi}\ and\ \citenamefont
  {Liakopoulos}(1995)}]{pinarbasi1995role}%
  \BibitemOpen
  \bibfield  {author} {\bibinfo {author} {\bibfnamefont {A.}~\bibnamefont
  {Pinarbasi}}\ and\ \bibinfo {author} {\bibfnamefont {A.}~\bibnamefont
  {Liakopoulos}},\ }\bibfield  {title} {\enquote {\bibinfo {title} {The role of
  variable viscosity in the stability of channel flow},}\ }\href@noop {}
  {\bibfield  {journal} {\bibinfo  {journal} {Int. Commun. Heat Mass Transf.}\
  }\textbf {\bibinfo {volume} {22}},\ \bibinfo {pages} {837--847} (\bibinfo
  {year} {1995})}\BibitemShut {NoStop}%
\bibitem [{\citenamefont {Wall}\ and\ \citenamefont
  {Wilson}(1996)}]{wall1996linear}%
  \BibitemOpen
  \bibfield  {author} {\bibinfo {author} {\bibfnamefont {D.~P.}\ \bibnamefont
  {Wall}}\ and\ \bibinfo {author} {\bibfnamefont {S.~K.}\ \bibnamefont
  {Wilson}},\ }\bibfield  {title} {\enquote {\bibinfo {title} {The linear
  stability of channel flow of fluid with temperature-dependent viscosity},}\
  }\href@noop {} {\bibfield  {journal} {\bibinfo  {journal} {J. Fluid Mech.}\
  }\textbf {\bibinfo {volume} {323}},\ \bibinfo {pages} {107--132} (\bibinfo
  {year} {1996})}\BibitemShut {NoStop}%
\bibitem [{\citenamefont {Sameen}\ and\ \citenamefont
  {Govindarajan}(2007)}]{sameen2007effect}%
  \BibitemOpen
  \bibfield  {author} {\bibinfo {author} {\bibfnamefont {A.}~\bibnamefont
  {Sameen}}\ and\ \bibinfo {author} {\bibfnamefont {R.}~\bibnamefont
  {Govindarajan}},\ }\bibfield  {title} {\enquote {\bibinfo {title} {The effect
  of wall heating on instability of channel flow},}\ }\href@noop {} {\bibfield
  {journal} {\bibinfo  {journal} {J. Fluid Mech.}\ }\textbf {\bibinfo {volume}
  {577}},\ \bibinfo {pages} {417--442} (\bibinfo {year} {2007})}\BibitemShut
  {NoStop}%
\bibitem [{\citenamefont {Sahu}\ and\ \citenamefont
  {Matar}(2010)}]{sahu2010stability}%
  \BibitemOpen
  \bibfield  {author} {\bibinfo {author} {\bibfnamefont {K.~C.}\ \bibnamefont
  {Sahu}}\ and\ \bibinfo {author} {\bibfnamefont {O.~K.}\ \bibnamefont
  {Matar}},\ }\bibfield  {title} {\enquote {\bibinfo {title} {Stability of
  plane channel flow with viscous heating},}\ }\href@noop {} {\bibfield
  {journal} {\bibinfo  {journal} {J. Fluids Eng.}\ }\textbf {\bibinfo {volume}
  {132}} (\bibinfo {year} {2010})}\BibitemShut {NoStop}%
\bibitem [{\citenamefont {Yueh}\ and\ \citenamefont
  {Weng}(1996)}]{yueh1996linear}%
  \BibitemOpen
  \bibfield  {author} {\bibinfo {author} {\bibfnamefont {C.~S.}\ \bibnamefont
  {Yueh}}\ and\ \bibinfo {author} {\bibfnamefont {C.~I.}\ \bibnamefont
  {Weng}},\ }\bibfield  {title} {\enquote {\bibinfo {title} {Linear stability
  analysis of plane {C}ouette flow with viscous heating},}\ }\href@noop {}
  {\bibfield  {journal} {\bibinfo  {journal} {Phys. Fluids}\ }\textbf {\bibinfo
  {volume} {8}},\ \bibinfo {pages} {1802--1813} (\bibinfo {year}
  {1996})}\BibitemShut {NoStop}%
\bibitem [{\citenamefont {Sukanek}, \citenamefont {Goldstein},\ and\
  \citenamefont {Laurence}(1973)}]{sukanek1973stability}%
  \BibitemOpen
  \bibfield  {author} {\bibinfo {author} {\bibfnamefont {P.~C.}\ \bibnamefont
  {Sukanek}}, \bibinfo {author} {\bibfnamefont {C.~A.}\ \bibnamefont
  {Goldstein}}, \ and\ \bibinfo {author} {\bibfnamefont {R.~L.}\ \bibnamefont
  {Laurence}},\ }\bibfield  {title} {\enquote {\bibinfo {title} {The stability
  of plane {Co}uette flow with viscous heating},}\ }\href@noop {} {\bibfield
  {journal} {\bibinfo  {journal} {J. Fluid Mech.}\ }\textbf {\bibinfo {volume}
  {57}},\ \bibinfo {pages} {651--670} (\bibinfo {year} {1973})}\BibitemShut
  {NoStop}%
\bibitem [{\citenamefont {Thomas}, \citenamefont {Sureshkumar},\ and\
  \citenamefont {Khomami}(2003)}]{thomas2003influence}%
  \BibitemOpen
  \bibfield  {author} {\bibinfo {author} {\bibfnamefont {D.~G.}\ \bibnamefont
  {Thomas}}, \bibinfo {author} {\bibfnamefont {R.}~\bibnamefont {Sureshkumar}},
  \ and\ \bibinfo {author} {\bibfnamefont {B.}~\bibnamefont {Khomami}},\
  }\bibfield  {title} {\enquote {\bibinfo {title} {Influence of fluid thermal
  sensitivity on the thermo-mechanical stability of the {T}aylor--{C}ouette
  flow},}\ }\href@noop {} {\bibfield  {journal} {\bibinfo  {journal} {Phys.
  Fluids}\ }\textbf {\bibinfo {volume} {15}},\ \bibinfo {pages} {3308--3317}
  (\bibinfo {year} {2003})}\BibitemShut {NoStop}%
\bibitem [{\citenamefont {Booker}(1976)}]{booker1976thermal}%
  \BibitemOpen
  \bibfield  {author} {\bibinfo {author} {\bibfnamefont {J.~R.}\ \bibnamefont
  {Booker}},\ }\bibfield  {title} {\enquote {\bibinfo {title} {Thermal
  convection with strongly temperature-dependent viscosity},}\ }\href@noop {}
  {\bibfield  {journal} {\bibinfo  {journal} {J. Fluid Mech.}\ }\textbf
  {\bibinfo {volume} {76}},\ \bibinfo {pages} {741--754} (\bibinfo {year}
  {1976})}\BibitemShut {NoStop}%
\bibitem [{\citenamefont {Booker}\ and\ \citenamefont
  {Stengel}(1978)}]{booker1978further}%
  \BibitemOpen
  \bibfield  {author} {\bibinfo {author} {\bibfnamefont {J.~R.}\ \bibnamefont
  {Booker}}\ and\ \bibinfo {author} {\bibfnamefont {K.~C.}\ \bibnamefont
  {Stengel}},\ }\bibfield  {title} {\enquote {\bibinfo {title} {Further
  thoughts on convective heat transport in a variable-viscosity fluid},}\
  }\href@noop {} {\bibfield  {journal} {\bibinfo  {journal} {J. Fluid Mech.}\
  }\textbf {\bibinfo {volume} {86}},\ \bibinfo {pages} {289--291} (\bibinfo
  {year} {1978})}\BibitemShut {NoStop}%
\bibitem [{\citenamefont {Stengel}, \citenamefont {Oliver},\ and\ \citenamefont
  {Booker}(1982)}]{stengel1982onset}%
  \BibitemOpen
  \bibfield  {author} {\bibinfo {author} {\bibfnamefont {K.~C.}\ \bibnamefont
  {Stengel}}, \bibinfo {author} {\bibfnamefont {D.~S.}\ \bibnamefont {Oliver}},
  \ and\ \bibinfo {author} {\bibfnamefont {J.~R.}\ \bibnamefont {Booker}},\
  }\bibfield  {title} {\enquote {\bibinfo {title} {Onset of convection in a
  variable-viscosity fluid},}\ }\href@noop {} {\bibfield  {journal} {\bibinfo
  {journal} {J. Fluid Mech.}\ }\textbf {\bibinfo {volume} {120}},\ \bibinfo
  {pages} {411--431} (\bibinfo {year} {1982})}\BibitemShut {NoStop}%
\bibitem [{\citenamefont {Thangam}\ and\ \citenamefont
  {Chen}(1986)}]{thangam1986stability}%
  \BibitemOpen
  \bibfield  {author} {\bibinfo {author} {\bibfnamefont {S.}~\bibnamefont
  {Thangam}}\ and\ \bibinfo {author} {\bibfnamefont {C.~F.}\ \bibnamefont
  {Chen}},\ }\bibfield  {title} {\enquote {\bibinfo {title} {Stability analysis
  on the convection of a variable viscosity fluid in an infinite vertical
  slot},}\ }\href@noop {} {\bibfield  {journal} {\bibinfo  {journal} {Phys.
  Fluids}\ }\textbf {\bibinfo {volume} {29}},\ \bibinfo {pages} {1367--1372}
  (\bibinfo {year} {1986})}\BibitemShut {NoStop}%
\bibitem [{\citenamefont {Ghosh}, \citenamefont {Usha},\ and\ \citenamefont
  {Sahu}(2014)}]{ghosh2014linear}%
  \BibitemOpen
  \bibfield  {author} {\bibinfo {author} {\bibfnamefont {S.}~\bibnamefont
  {Ghosh}}, \bibinfo {author} {\bibfnamefont {R.}~\bibnamefont {Usha}}, \ and\
  \bibinfo {author} {\bibfnamefont {K.~C.}\ \bibnamefont {Sahu}},\ }\bibfield
  {title} {\enquote {\bibinfo {title} {Linear stability analysis of miscible
  two-fluid flow in a channel with velocity slip at the walls},}\ }\href@noop
  {} {\bibfield  {journal} {\bibinfo  {journal} {Phys. Fluids}\ }\textbf
  {\bibinfo {volume} {26}},\ \bibinfo {pages} {014107} (\bibinfo {year}
  {2014})}\BibitemShut {NoStop}%
\bibitem [{\citenamefont {Chattopadhyay}, \citenamefont {Usha},\ and\
  \citenamefont {Sahu}(2017)}]{chattopadhyay2017core}%
  \BibitemOpen
  \bibfield  {author} {\bibinfo {author} {\bibfnamefont {G.}~\bibnamefont
  {Chattopadhyay}}, \bibinfo {author} {\bibfnamefont {R.}~\bibnamefont {Usha}},
  \ and\ \bibinfo {author} {\bibfnamefont {K.~C.}\ \bibnamefont {Sahu}},\
  }\bibfield  {title} {\enquote {\bibinfo {title} {Core-annular miscible
  two-fluid flow in a slippery pipe: A stability analysis},}\ }\href@noop {}
  {\bibfield  {journal} {\bibinfo  {journal} {Phys. Fluids}\ }\textbf {\bibinfo
  {volume} {29}},\ \bibinfo {pages} {097106} (\bibinfo {year}
  {2017})}\BibitemShut {NoStop}%
\bibitem [{\citenamefont {Pramanik}\ and\ \citenamefont
  {Mishra}(2013)}]{pramanik2013linear}%
  \BibitemOpen
  \bibfield  {author} {\bibinfo {author} {\bibfnamefont {S.}~\bibnamefont
  {Pramanik}}\ and\ \bibinfo {author} {\bibfnamefont {M.}~\bibnamefont
  {Mishra}},\ }\bibfield  {title} {\enquote {\bibinfo {title} {Linear stability
  analysis of {K}orteweg stresses effect on miscible viscous fingering in
  porous media},}\ }\href@noop {} {\bibfield  {journal} {\bibinfo  {journal}
  {Phys. Fluids}\ }\textbf {\bibinfo {volume} {25}},\ \bibinfo {pages} {074104}
  (\bibinfo {year} {2013})}\BibitemShut {NoStop}%
\bibitem [{\citenamefont {Ranganathan}\ and\ \citenamefont
  {Govindarajan}(2001)}]{ranganathan2001stabilization}%
  \BibitemOpen
  \bibfield  {author} {\bibinfo {author} {\bibfnamefont {B.~T.}\ \bibnamefont
  {Ranganathan}}\ and\ \bibinfo {author} {\bibfnamefont {R.}~\bibnamefont
  {Govindarajan}},\ }\bibfield  {title} {\enquote {\bibinfo {title}
  {Stabilization and destabilization of channel flow by location of
  viscosity-stratified fluid layer},}\ }\href@noop {} {\bibfield  {journal}
  {\bibinfo  {journal} {Phys. Fluids}\ }\textbf {\bibinfo {volume} {13}},\
  \bibinfo {pages} {1--3} (\bibinfo {year} {2001})}\BibitemShut {NoStop}%
\bibitem [{\citenamefont {Govindarajan}(2004)}]{govindarajan2004effect}%
  \BibitemOpen
  \bibfield  {author} {\bibinfo {author} {\bibfnamefont {R.}~\bibnamefont
  {Govindarajan}},\ }\bibfield  {title} {\enquote {\bibinfo {title} {Effect of
  miscibility on the linear instability of two-fluid channel flow},}\
  }\href@noop {} {\bibfield  {journal} {\bibinfo  {journal} {Int. J. Multiph.
  Flow}\ }\textbf {\bibinfo {volume} {30}},\ \bibinfo {pages} {1177--1192}
  (\bibinfo {year} {2004})}\BibitemShut {NoStop}%
\bibitem [{\citenamefont {Turner}(1974)}]{turner1974double}%
  \BibitemOpen
  \bibfield  {author} {\bibinfo {author} {\bibfnamefont {J.~S.}\ \bibnamefont
  {Turner}},\ }\bibfield  {title} {\enquote {\bibinfo {title} {Double-diffusive
  phenomena},}\ }\href@noop {} {\bibfield  {journal} {\bibinfo  {journal}
  {Annu. Rev. Fluid Mech.}\ }\textbf {\bibinfo {volume} {6}},\ \bibinfo {pages}
  {37--54} (\bibinfo {year} {1974})}\BibitemShut {NoStop}%
\bibitem [{\citenamefont {Sahu}(2014)}]{sahu2014review}%
  \BibitemOpen
  \bibfield  {author} {\bibinfo {author} {\bibfnamefont {K.~C.}\ \bibnamefont
  {Sahu}},\ }\bibfield  {title} {\enquote {\bibinfo {title} {A review on
  double-diffusive instability in viscosity stratified flows},}\ }in\
  \href@noop {} {\emph {\bibinfo {booktitle} {Proc Indian Natn Sci Acad}}},\
  Vol.~\bibinfo {volume} {80}\ (\bibinfo {organization} {Citeseer},\ \bibinfo
  {year} {2014})\ pp.\ \bibinfo {pages} {513--514}\BibitemShut {NoStop}%
\bibitem [{\citenamefont {Sahu}(2020)}]{sahu2020linear}%
  \BibitemOpen
  \bibfield  {author} {\bibinfo {author} {\bibfnamefont {K.~C.}\ \bibnamefont
  {Sahu}},\ }\bibfield  {title} {\enquote {\bibinfo {title} {Linear instability
  in two-layer channel flow due to double-diffusive phenomenon},}\ }\href@noop
  {} {\bibfield  {journal} {\bibinfo  {journal} {Phys. Fluids}\ }\textbf
  {\bibinfo {volume} {32}},\ \bibinfo {pages} {024102} (\bibinfo {year}
  {2020})}\BibitemShut {NoStop}%
\bibitem [{\citenamefont {Sahu}\ and\ \citenamefont
  {Govindarajan}(2011)}]{sahu2011linear}%
  \BibitemOpen
  \bibfield  {author} {\bibinfo {author} {\bibfnamefont {K.~C.}\ \bibnamefont
  {Sahu}}\ and\ \bibinfo {author} {\bibfnamefont {R.}~\bibnamefont
  {Govindarajan}},\ }\bibfield  {title} {\enquote {\bibinfo {title} {Linear
  stability of double-diffusive two-fluid channel flow},}\ }\href@noop {}
  {\bibfield  {journal} {\bibinfo  {journal} {J. Fluid Mech.}\ }\textbf
  {\bibinfo {volume} {687}},\ \bibinfo {pages} {529--539} (\bibinfo {year}
  {2011})}\BibitemShut {NoStop}%
\bibitem [{\citenamefont {Sahu}\ and\ \citenamefont
  {Govindarajan}(2012)}]{sahu2012spatio}%
  \BibitemOpen
  \bibfield  {author} {\bibinfo {author} {\bibfnamefont {K.~C.}\ \bibnamefont
  {Sahu}}\ and\ \bibinfo {author} {\bibfnamefont {R.}~\bibnamefont
  {Govindarajan}},\ }\bibfield  {title} {\enquote {\bibinfo {title}
  {Spatio-temporal linear stability of double-diffusive two-fluid channel
  flow},}\ }\href@noop {} {\bibfield  {journal} {\bibinfo  {journal} {Phys.
  Fluids}\ }\textbf {\bibinfo {volume} {24}},\ \bibinfo {pages} {054103}
  (\bibinfo {year} {2012})}\BibitemShut {NoStop}%
\bibitem [{\citenamefont {Swernath}\ and\ \citenamefont
  {Pushpavanam}(2007)}]{swernath2007viscous}%
  \BibitemOpen
  \bibfield  {author} {\bibinfo {author} {\bibfnamefont {S.}~\bibnamefont
  {Swernath}}\ and\ \bibinfo {author} {\bibfnamefont {S.}~\bibnamefont
  {Pushpavanam}},\ }\bibfield  {title} {\enquote {\bibinfo {title} {Viscous
  fingering in a horizontal flow through a porous medium induced by chemical
  reactions under isothermal and adiabatic conditions},}\ }\href@noop {}
  {\bibfield  {journal} {\bibinfo  {journal} {J. Chem. Phys.}\ }\textbf
  {\bibinfo {volume} {127}},\ \bibinfo {pages} {204701} (\bibinfo {year}
  {2007})}\BibitemShut {NoStop}%
\bibitem [{\citenamefont {Mishra}\ \emph {et~al.}(2010)\citenamefont {Mishra},
  \citenamefont {Trevelyan}, \citenamefont {Almarcha},\ and\ \citenamefont
  {Wit}}]{mishra2010influence}%
  \BibitemOpen
  \bibfield  {author} {\bibinfo {author} {\bibfnamefont {M.}~\bibnamefont
  {Mishra}}, \bibinfo {author} {\bibfnamefont {P.~M.~J.}\ \bibnamefont
  {Trevelyan}}, \bibinfo {author} {\bibfnamefont {C.}~\bibnamefont {Almarcha}},
  \ and\ \bibinfo {author} {\bibfnamefont {A.~D.}\ \bibnamefont {Wit}},\
  }\bibfield  {title} {\enquote {\bibinfo {title} {Influence of double
  diffusive effects on miscible viscous fingering},}\ }\href@noop {} {\bibfield
   {journal} {\bibinfo  {journal} {Phys. Rev. Lett.}\ }\textbf {\bibinfo
  {volume} {105}},\ \bibinfo {pages} {204501} (\bibinfo {year}
  {2010})}\BibitemShut {NoStop}%
\bibitem [{\citenamefont {Pritchard}(2009)}]{pritchard2009linear}%
  \BibitemOpen
  \bibfield  {author} {\bibinfo {author} {\bibfnamefont {D.}~\bibnamefont
  {Pritchard}},\ }\bibfield  {title} {\enquote {\bibinfo {title} {The linear
  stability of double-diffusive miscible rectilinear displacements in a
  {H}ele-{S}haw cell},}\ }\href@noop {} {\bibfield  {journal} {\bibinfo
  {journal} {Eur. J. Mech. B/Fluids}\ }\textbf {\bibinfo {volume} {28}},\
  \bibinfo {pages} {564--577} (\bibinfo {year} {2009})}\BibitemShut {NoStop}%
\bibitem [{\citenamefont {Bratsun}\ \emph {et~al.}(2022)\citenamefont
  {Bratsun}, \citenamefont {Oschepkov}, \citenamefont {Mosheva},\ and\
  \citenamefont {Siraev}}]{bratsun2022effect}%
  \BibitemOpen
  \bibfield  {author} {\bibinfo {author} {\bibfnamefont {D.~A.}\ \bibnamefont
  {Bratsun}}, \bibinfo {author} {\bibfnamefont {V.~O.}\ \bibnamefont
  {Oschepkov}}, \bibinfo {author} {\bibfnamefont {E.~A.}\ \bibnamefont
  {Mosheva}}, \ and\ \bibinfo {author} {\bibfnamefont {R.~R.}\ \bibnamefont
  {Siraev}},\ }\bibfield  {title} {\enquote {\bibinfo {title} {The effect of
  concentration-dependent diffusion on double-diffusive instability},}\
  }\href@noop {} {\bibfield  {journal} {\bibinfo  {journal} {Phys. Fluids}\
  }\textbf {\bibinfo {volume} {34}},\ \bibinfo {pages} {034112} (\bibinfo
  {year} {2022})}\BibitemShut {NoStop}%
\bibitem [{\citenamefont {Mishra}, \citenamefont {Wit},\ and\ \citenamefont
  {Sahu}(2012)}]{mishra2012double}%
  \BibitemOpen
  \bibfield  {author} {\bibinfo {author} {\bibfnamefont {M.}~\bibnamefont
  {Mishra}}, \bibinfo {author} {\bibfnamefont {A.~D.}\ \bibnamefont {Wit}}, \
  and\ \bibinfo {author} {\bibfnamefont {K.~C.}\ \bibnamefont {Sahu}},\
  }\bibfield  {title} {\enquote {\bibinfo {title} {Double diffusive effects on
  pressure-driven miscible displacement flows in a channel},}\ }\href@noop {}
  {\bibfield  {journal} {\bibinfo  {journal} {J. Fluid Mech.}\ }\textbf
  {\bibinfo {volume} {712}},\ \bibinfo {pages} {579--597} (\bibinfo {year}
  {2012})}\BibitemShut {NoStop}%
\bibitem [{\citenamefont {Verma}, \citenamefont {Sharma},\ and\ \citenamefont
  {Mishra}(2022)}]{verma2022radial}%
  \BibitemOpen
  \bibfield  {author} {\bibinfo {author} {\bibfnamefont {P.}~\bibnamefont
  {Verma}}, \bibinfo {author} {\bibfnamefont {V.}~\bibnamefont {Sharma}}, \
  and\ \bibinfo {author} {\bibfnamefont {M.}~\bibnamefont {Mishra}},\
  }\bibfield  {title} {\enquote {\bibinfo {title} {Radial viscous fingering
  induced by an infinitely fast chemical reaction},}\ }\href@noop {} {\bibfield
   {journal} {\bibinfo  {journal} {J. Fluid Mech.}\ }\textbf {\bibinfo {volume}
  {945}},\ \bibinfo {pages} {A19} (\bibinfo {year} {2022})}\BibitemShut
  {NoStop}%
\bibitem [{\citenamefont {Maharana}\ and\ \citenamefont
  {Mishra}(2022)}]{maharana2022effects}%
  \BibitemOpen
  \bibfield  {author} {\bibinfo {author} {\bibfnamefont {S.~N.}\ \bibnamefont
  {Maharana}}\ and\ \bibinfo {author} {\bibfnamefont {M.}~\bibnamefont
  {Mishra}},\ }\bibfield  {title} {\enquote {\bibinfo {title} {Effects of low
  and high viscous product on {K}elvin--{H}elmholtz instability triggered by
  {A+ B $\rightarrow$ C} type reaction},}\ }\href@noop {} {\bibfield  {journal}
  {\bibinfo  {journal} {Phys. Fluids}\ }\textbf {\bibinfo {volume} {34}},\
  \bibinfo {pages} {012104} (\bibinfo {year} {2022})}\BibitemShut {NoStop}%
\bibitem [{\citenamefont {Maharana}, \citenamefont {Sahu},\ and\ \citenamefont
  {Mishra}(2023)}]{maharana2023stability}%
  \BibitemOpen
  \bibfield  {author} {\bibinfo {author} {\bibfnamefont {S.~N.}\ \bibnamefont
  {Maharana}}, \bibinfo {author} {\bibfnamefont {K.~C.}\ \bibnamefont {Sahu}},
  \ and\ \bibinfo {author} {\bibfnamefont {M.}~\bibnamefont {Mishra}},\
  }\bibfield  {title} {\enquote {\bibinfo {title} {Stability of a layered
  reactive channel flow},}\ }\href@noop {} {\bibfield  {journal} {\bibinfo
  {journal} {Proc. R. Soc. A}\ }\textbf {\bibinfo {volume} {479}},\ \bibinfo
  {pages} {20220689} (\bibinfo {year} {2023})}\BibitemShut {NoStop}%
\bibitem [{\citenamefont {Khandelwal}\ \emph {et~al.}(2021)\citenamefont
  {Khandelwal}, \citenamefont {Singh}, \citenamefont {Sharma},\ and\
  \citenamefont {Yu}}]{khandelwal2021instabilities}%
  \BibitemOpen
  \bibfield  {author} {\bibinfo {author} {\bibfnamefont {M.~K.}\ \bibnamefont
  {Khandelwal}}, \bibinfo {author} {\bibfnamefont {N.}~\bibnamefont {Singh}},
  \bibinfo {author} {\bibfnamefont {A.~K.}\ \bibnamefont {Sharma}}, \ and\
  \bibinfo {author} {\bibfnamefont {P.}~\bibnamefont {Yu}},\ }\bibfield
  {title} {\enquote {\bibinfo {title} {Instabilities during
  convection--diffusion of binary mixtures in a non-isothermal flow: A linear
  stability analysis},}\ }\href@noop {} {\bibfield  {journal} {\bibinfo
  {journal} {Phys. Fluids}\ }\textbf {\bibinfo {volume} {33}},\ \bibinfo
  {pages} {084107} (\bibinfo {year} {2021})}\BibitemShut {NoStop}%
\bibitem [{\citenamefont {Nahme}(1940)}]{nahme1940beitrage}%
  \BibitemOpen
  \bibfield  {author} {\bibinfo {author} {\bibfnamefont {R.}~\bibnamefont
  {Nahme}},\ }\bibfield  {title} {\enquote {\bibinfo {title} {Beitr{\"a}ge zur
  hydrodynamischen theorie der lagerreibung},}\ }\href@noop {} {\bibfield
  {journal} {\bibinfo  {journal} {Ing.-Arch.}\ }\textbf {\bibinfo {volume}
  {11}},\ \bibinfo {pages} {191--209} (\bibinfo {year} {1940})}\BibitemShut
  {NoStop}%
\bibitem [{\citenamefont {Rayleigh}(1879)}]{rayleigh1879stability}%
  \BibitemOpen
  \bibfield  {author} {\bibinfo {author} {\bibfnamefont {L.}~\bibnamefont
  {Rayleigh}},\ }\bibfield  {title} {\enquote {\bibinfo {title} {On the
  stability, or instability, of certain fluid motions},}\ }\href@noop {}
  {\bibfield  {journal} {\bibinfo  {journal} {Proceedings of the London
  Mathematical Society}\ }\textbf {\bibinfo {volume} {1}},\ \bibinfo {pages}
  {57--72} (\bibinfo {year} {1879})}\BibitemShut {NoStop}%
\bibitem [{\citenamefont {Canuto}\ \emph {et~al.}(1988)\citenamefont {Canuto},
  \citenamefont {Hussaini}, \citenamefont {Quarteroni},\ and\ \citenamefont
  {Zang}}]{canuto1988spectral}%
  \BibitemOpen
  \bibfield  {author} {\bibinfo {author} {\bibfnamefont {C.}~\bibnamefont
  {Canuto}}, \bibinfo {author} {\bibfnamefont {M.~Y.}\ \bibnamefont
  {Hussaini}}, \bibinfo {author} {\bibfnamefont {A.}~\bibnamefont
  {Quarteroni}}, \ and\ \bibinfo {author} {\bibfnamefont {T.~A.}\ \bibnamefont
  {Zang}},\ }\href@noop {} {\emph {\bibinfo {title} {Spectral Method in Fluid
  Dynamics}}}\ (\bibinfo  {publisher} {Springer, New York, Berlin,
  Heidelberg},\ \bibinfo {year} {1988})\BibitemShut {NoStop}%
\end{thebibliography}

%

\end{document}